%% file: directionality_matters_arxiv.tex
\documentclass[letterpaper,twocolumn,10pt]{article}
\usepackage{usenix-2020-09}

\usepackage{amsmath}
\usepackage{numprint}
\usepackage{standalone}
\usepackage{xcolor}
\usepackage{siunitx}
\usepackage{tikz}
\usepackage{layouts}
\usepackage{mdframed}
\usepackage{xurl}
\usepackage{booktabs}
\usepackage{makecell}
\usepackage{multirow}
\usepackage{subcaption}
\usepackage{mfirstuc}	%
\usetikzlibrary{positioning, shapes, calc, arrows.meta, matrix, backgrounds}
\definecolor{perception_background}{HTML}{A8B8AE}
\definecolor{technology_background}{HTML}{C8C6AD}
\definecolor{acquisition_background}{HTML}{FDDDD1}
\definecolor{gray}{HTML}{D8D8D8}
\tikzset{>=stealth}

\definecolor{green}{RGB}{153, 204, 0}
\definecolor{red}{RGB}{204, 0, 0}
\definecolor{textcolor}{RGB}{62,62,59}
\definecolor{maincolor}{cmyk}{0,0.5,1,0} %
\definecolor{altcolor}{cmyk}{1,0.6,0,0.56} %
\colorlet{mainlight}{maincolor!50}
\colorlet{altlight}{altcolor!50}
\colorlet{textlight}{textcolor!70}
\colorlet{alert}{red}

\graphicspath{{figures/}}

\mdfsetup{skipabove=12pt,skipbelow=8pt}

\mdfdefinestyle{observation}{%
	linecolor=black,
	outerlinewidth=2pt,
	backgroundcolor=black!10,
}

\newcommand{\libjpeg}{\emph{libjpeg}}
\newcommand{\turbo}{\emph{libjpeg-turbo}}
\newcommand{\mozjpeg}{\emph{MozJPEG}}  %

\begin{document}
\date{}

\title{\Large \bf Landscape More Secure Than Portrait\,?\\Zooming Into the Directionality of Digital Images With Security Implications}
\author{
{\rm Benedikt Lorch}\\
University of Innsbruck
\and
{\rm Rainer Böhme}\\
University of Innsbruck
} %

\maketitle

\begin{abstract}

The orientation in which a source image is captured can affect the resulting security in downstream applications.
One reason for this is that many state-of-the-art methods in media security assume that image statistics are similar in the horizontal and vertical directions, allowing them to reduce the number of features (or trainable weights) by merging coefficients. 
We show that this artificial symmetrization tends to suppress important properties of natural images and common processing operations, causing a loss of performance.
We also observe the opposite problem, where unaddressed directionality causes learning-based methods to overfit to a single orientation. 
These are vulnerable to manipulation if an adversary chooses inputs with the less common orientation.
This paper takes a comprehensive approach, identifies and systematizes causes of directionality at several stages of a typical acquisition pipeline, measures their effect, and demonstrates for three selected security applications (steganalysis, forensic source identification, and the detection of synthetic images) how the performance of state-of-the-art methods can be improved by properly accounting for directionality.

\end{abstract}

\section{Introduction}
\label{sec:introduction}

Images are \emph{directional} 
if horizontal pixels sequences have different statistical properties than vertical sequences.
Directionality has multiple causes. 
It can originate from scene content, unintentional asymmetries in image processing, or technological choices that appeal to human visual perception. 
While the effect of directionality has so far been neglected in security applications that depend on images, we show that directionality impacts the security of image steganography, forensic source identification, and the detection of synthetic images (aka ``fake-or-real'').

\begin{figure}

	\begin{center}
	\begin{tikzpicture}[x=27mm,y=70mm]
		\draw (-2.75em,.8) node [rotate=90] {Accuracy};

		\draw (.3,1.05) node [above] {\parbox{.3\linewidth}{\centering Steganography detection}};

		\draw (0,.6) rectangle (.6,1.03);

		\foreach \y in {0.6,0.7,0.8,0.9,1.0}
			\draw (0,\y)--++(-4pt,0) node [left] {\small \y};

		\begin{scope}
			\clip (0,.6) rectangle (.6,1);		
			\draw [altcolor,line width=6mm] (.15,0)--++(0,.911);
			\draw [maincolor,line width=6mm] (.45,0)--++(0,.780);
		\end{scope}

		\draw (.15,.6)--++(0,-4pt) node [below] {orig.\strut};
		\draw (.45,.6)--++(0,-4pt) node [below] {rot.\strut};

		\draw (1.3,1.05) node [above] {\parbox{.3\linewidth}{\centering Forensic source identification}};
		\draw (1,.6) coordinate (A) rectangle (1.6,1.03);
		
		\begin{scope}[yscale=1.34,yshift=-17.65mm]
			\foreach \y in {0.7,0.8,0.9,1.0}
				\draw (1,\y)--++(-4pt,0) node [left] {\small \y};

			\begin{scope}
				\clip (A) rectangle (2.6,1);		
				\draw [altcolor,line width=6mm] (1.15,0)--++(0,.941);
				\draw [maincolor,line width=6mm] (1.45,0)--++(0,.845);
			\end{scope}
		\end{scope}

		\draw (1.15,.6)--++(0,-4pt) node [below] {orig.\strut};
		\draw (1.45,.6)--++(0,-4pt) node [below] {rot.\strut};

		\draw (2.3,1.05) node [above] {\parbox{.3\linewidth}{\centering Image authentication}};
		\draw (2,.6) rectangle (2.6,1.03);
		\foreach \y in {0.6,0.7,0.8,0.9,1.0}
			\draw (2,\y)--++(-4pt,0) node [left] {\small\y};
				
		\begin{scope}
			\clip (2,.6) rectangle (2.6,1);		
			\draw [altcolor,line width=6mm] (2.15,0)--++(0,1);
			\draw [maincolor,line width=6mm] (2.45,0)--++(0,.674);
		\end{scope}

		\draw (2.15,.6)--++(0,-4pt) node [below] {orig.\strut};
		\draw (2.45,.6)--++(0,-4pt) node [below] {rot.\strut};
		
	\end{tikzpicture}
	\end{center}
	\vspace{-3ex}
	\caption{Directionality matters across security applications. The figure summarizes the performance loss of selected state-of-the-art classification tasks when the (square-shaped) test images are rotated by 90 degrees. Note the different scales.} 
	\label{fig:introduction}
\end{figure}
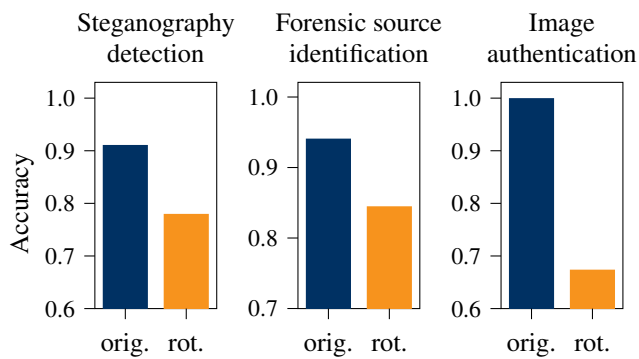

Figure~\ref{fig:introduction} showcases the loss of not considering directionality in these three applications.
The task of \emph{steganalysis} is to attack steganography~\cite{Simmons1984Prisoner} by detecting hidden messages embedded in innocuous cover images~\cite{Provos2001OutGuess}.
We embed 0.4 bit per usable sample into JPEG images of size $512^2$ using J-UNIWARD~\cite{Holub2014Uniward}, a state-of-the art steganography method, resulting in an average payload ratio of 4.3\,\%.
The leftmost graph shows the detection accuracy of an EfficientNet-B0~\cite{Tan2019EfficientNet} detector with 4\,M parameters trained on the dataset from the ALASKA2 steganalysis competition~\cite{Cogranne2020Alaska2}. 
It achieves 91\,\% accuracy (blue bar) for the binary task of distinguishing between stego images and pristine covers.
This accuracy drops to 78\,\% (orange bar) when the test images are rotated by 90 degrees, indicating a lack of robustness to directionality.
In terms of error rates, this means that the probability of false accusations increases from 8\,\% to 37\,\%.
(See Sec.~\ref{sec:steganalysis} for more details.)
A common task in \emph{forensic source identification} is to identify the camera model that was used to acquire a given image~\cite{Kirchner2015CameraModel, Mandelli2022CameraModelIdentification}.
The center graph shows the accuracy of EfficientNet-B5 trained as multi-class classifier on scenes from the Forchheim database~\cite{Hadwiger2020Forchheim}, which contains images acquired with 27 different smartphone models.
Again, the performance drops from 94\,\% to 85\,\% when the square-shaped test images are rotated by 90 degrees.
(See Sec.~\ref{sec:identification} for details.)
Finally, \emph{image authentication} concerns the task of distinguishing real images from forgeries~\cite{Farid2016PhotoForensics}.
An increasingly important subtask is the detection of synthetic images, which has received considerable attention through large-scale challenges including ones hosted by Facebook~\cite{Dolhansky2020DeepFakeDetectionChallenge} and the IEEE~\cite{Cozzolino2023VipCup}.
The rightmost graph shows the performance of EfficientNet-B0 trained on the binary classification task of distinguishing between real images from the ALASKA2 dataset and generated images using the Stable Diffusion XL~\cite{Podell2023StableDiffusionXL} autoencoder.
Once again, the accuracy drops significantly from 100\,\% to 67\,\% if the orientation of the test images does not match the training images.
(See Sec.~\ref{sec:synthetic} for details.)
In all examples, we replicate a common experimental setup and train a state-of-the-art CNN detector.
The results reveal that images contain directional statistics, which learning-based detectors tend to exploit. 
This calls for a spotlight on the effect of directionality in security applications.

To give one example for a typical cause of directional artifacts, consider JPEG compression~\cite{Pennebaker1992JPEGStandard}. %
While this popular~\cite{Hudson2018Jpeg25Years, Dornauer2023Formats} lossy compression standard does not favor any direction, common parameter choices do leave directional artifacts. 
These include the practice to subsample the chroma channels in the horizontal direction only, or the asymmetry in the default quantization table.
Some practices can be explained with technological requirements (e.g., interlacing on CRT monitors~\cite{Jancovic2023Interlacing}); others originate from empirical measurements of human visual perception (e.g., contrast sensitivity by spatial frequency~\cite{Lohscheller1984Visibility}). %
Strikingly, these historically justified practices prevail in contemporary cameras and software libraries.
Multiple causes of directionality may coexist, interact with directional biases in the scene content, and therefore require careful analysis. 
There is no single measure of directionality and causal relationships are not always straightforward.

\paragraph{Contributions}

This paper takes a comprehensive approach to studying directionality of images in security applications. 
It makes the following specific contributions:
\begin{itemize}
	 \setlength\itemsep{.5ex}
	\item We identify and systematize causes of directionality and relate them to technology and human perception.
	\item We present evidence for the effect of directionality in selected security applications: steganalysis, forensic source identification, and the detection of synthetic images.
	\item We show how the state of the art in each application can be improved by correctly accounting for directionality.
	\item We extract guidance on the use of rotation augmentation, which has the potential to generalize across applications.  %
\end{itemize}

\paragraph{Implications}
Our work has at least three important implications.
First and foremost, if security techniques are biased towards one direction, then the opponent can have an advantage by choosing images of the opposite orientation.
Second, defenders who want to consider directionality face the decision to incur higher cost and complexity (e.g., in terms of trainable weights) or accept limited generalizability across image orientations.
This applies in particular to deep learning-based methods, but also to hand-crafted features (as we show in Sec.~\ref{sec:identification}).
Lastly, some causes we identify are due to legacy decisions made for technology that is no longer in use (e.g., CRT monitors). 
Standardization bodies and developers of image processing libraries should revisit these conventions.

\paragraph{Organization}
Section~\ref{sec:measuring} discusses how directionality can be measured. 
Section~\ref{sec:causes} systematizes the causes of directionality and relates them to technology and human perception. 
The three following sections evaluate the effect on selected security applications: steganalysis (Sec.~\ref{sec:steganalysis}), forensic camera model identification (Sec.~\ref{sec:identification}), and synthetic image detection (Sec.~\ref{sec:synthetic}). 
Section~\ref{sec:related} reviews related work. 
Section~\ref{sec:discussion} discusses implications and limitations, and Sec.~\ref{sec:conclusion} concludes.

\section{Measuring directionality}
\label{sec:measuring}

Tools to detect directional artifacts in images and, where possible, to quantify their strength are a prerequisite for evaluating the effects of directionality.
Directional artifacts occur at different stages of the image acquisition pipeline and manifest themselves in different statistical properties. 
To deal with this variety, we use several methods to measure directionality.
We discuss them below, from coarse to fine.

\paragraph{End-to-end effect}
	The three security applications mentioned in the introduction use a learning-based detector with measurable accuracy. 
	An intuitive measure of directionality is to evaluate the detector's sensitivity to rotated test images. 
	If the images were non-directional or directionality did not matter, the detector would naturally generalize to rotated test images. 
	A significant performance drop indicates that the detector is sensitive to directional image statistics, and the magnitude of the drop can be interpreted as effect size.
	
	An advantage of the end-to-end approach is that it measures the effect of directionality for the target application. 
	As learning-based detectors tend to exploit small nuances in image statistics, this approach can pick up subtle directional artifacts, which are difficult to measure directly with other tools.
	At the same time, the indirect measurement through a specific detector has several limitations. 
	A drop in performance is hardly comparable across applications.
	Moreover, this approach cannot reveal the cause of directionality.
	
	\begin{figure}
		\tikz{%
				\node [inner sep=0] (A) {\includegraphics{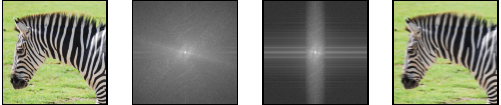}};
				\draw (A.39) node [below,white] {\scriptsize $d=-1.0$\strut};
				\draw (A.141) node [below,white] {\scriptsize $d=1.4$\strut};
		}
		\caption{The power spectra (center panels) reveal that the right image has low energy in the high horizontal frequencies. (Motion blur has been applied for illustration.)}
		\label{fig:power_spectrum_horiontal_blur}
	\end{figure}

\paragraph{Visualization}
	Statistical patterns in single images are often difficult to see with bare eyes. 
	A common technique in the forensics community is to visually inspect the power spectrum, where periodic patterns manifest themselves as peaks in the frequency domain~\cite{Popescu2005Resampling}.
	Figure~\ref{fig:power_spectrum_horiontal_blur} shows an example image (left) and its power spectrum (center left). Horizontal motion blur (right) removes high horizontal frequencies, as can be seen in the corresponding power spectrum (center right).
	
	Directional artifacts that are common and synchronous in all images from a particular source or processing operation can be amplified by averaging many images with diverse content.
	To further suppress the scene content, one can extract a noise residual.
	The noise residual is obtained by subtracting the image from a denoised version of it. 
	We use the denoising CNN (DnDNN)~\cite{Zhang2017DnCNN} with pre-trained weights from~\cite{Corvi2023DiffusionFingerprints}.
	Denoising and averaging can also be combined.
	
	Visualization has the advantage that directionality can be visually perceived and its orientation can be determined.
	Limitations are that subtle artifacts may not be perceptible, and that it is not easy to quantify the strength of directionality.
	
	\begin{figure}
		\includegraphics{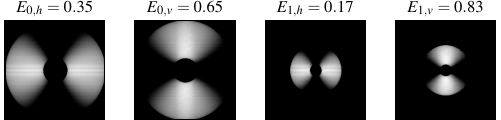}
		\caption{The steerable pyramids directionality score compares the horizontal and vertical energy on two scales.}
		\label{fig:steerable_pyramids}
	\end{figure}

\paragraph{Quantification by steerable pyramids}
	Ideally, we seek to quantify the directionality of an individual image.
	A numeric directionality score should have three properties:
	(1) a score of zero means that directional statistics are unchanged when the image orientation is changed; %
	(2) the sign of the score should indicate the dominant image direction; and 
	(3) the absolute value of the score should reflect the strength of directionality.
	
	The \emph{steerable pyramids} method~\cite{Simoncelli1995SteerablePyramids} has these properties. 
	It decomposes the image into several oriented frequency bands at different scales, as shown in Fig.~\ref{fig:steerable_pyramids}.
	For each scale $s$, it calculates the difference in energy between the two horizontal wedges $E_{s, h}$ and the two vertical wedges $E_{s, v}$, normalized by their sum. 
	The resulting scores are summed over all scales.
	Using two scales gives a directionality score in the range $[-2, 2]$.
	A positive score indicates that the image contains more energy in its horizontal frequencies than in its vertical frequencies. 
	Note that summing over the scales may cancel out directional differences at different scales. 
	Summing over absolute values would avoid this problem, but loses the sign of the dominant image direction. 

	We also experimented with alternative measurements. 
	An intuitive approach is to compute directional edges using first-order derivative filters, such as the Sobel operator~\cite{Sobel2014GradientOperator}.
	A directionality score can be defined as the difference between the horizontal and vertical absolute-valued derivative image. 
	We found that both detectors can indicate the dominant image direction, but the steerable pyramids are more precise. 
	The comparison is reported in Appendix~\ref{app:measuring_directionality}.
	
	Steerable pyramids are applicable to individual images and produce a numerical score $d$, which can be compared to other measurements.
	For example, the left image in Fig.~\ref{fig:power_spectrum_horiontal_blur} has a directionality score of $d = 1.4$.
	The prevalence of horizontal frequencies corresponds to the Zebra's vertical stripes.
	The right image has a score of $d = -1.0$, because the horizontal blur attenuates the horizontal frequencies and adds vertical frequencies.
	A limitation is that the method aggregates over the entire spectrum and is therefore less sensitive to directional artifacts in narrow subbands.
	By design, it cannot detect higher-order dependencies with directionality, e.g., pairwise correlation between horizontally (but not vertically) adjacent pixels that only occurs in certain environments.
	Only the learning-based detectors of the end-to-end approach are able to exploit this.
	This justifies our combination of approaches.

\section{Causes of directionality}
\label{sec:causes}

	\begin{figure*}
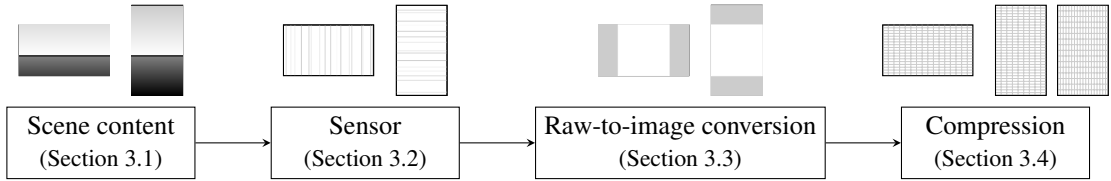

		\centering
		\includestandalone{figures/acquisition_pipeline}
		\caption{Causes of directionality are present in several stages of the image acquisition.}
		\label{fig:acqusition_pipeline}
	\end{figure*}

	Directionality in images has multiple causes. 
	It can be inherent to the image content or introduced at several stages during the image acquisition, as illustrated in Fig.~\ref{fig:acqusition_pipeline}.
	During image acquisition, the scene is projected through lenses onto a sensor. 
	The sensor turns the incoming light into an electrical current, which is amplified and converted to a raw digital signal. 
	The signal then undergoes a number of raw-to-image conversion steps including demosaicing, denoising, sharpening, color adjustments, gamma correction, quantization, and optionally lens correction.  
	The final image is usually stored in a compressed format, with JPEG still being most popular~\cite{Hudson2017Jpeg25, Park2018QuantizationTables, Dornauer2023Formats}. %
	All stages can introduce directional artifacts. 
	
	The following subsections walk through these stages and explain causes of directionality. 
	Each subsection starts with an observation in a gray box, then presents links to technology, human perception, and prior evidence. 
	At the end of each subsection, we present our own evidence in support of hypotheses derived from previous work.
	Appendix~\ref{app:data} gives an overview of the image sources used for the validation experiments.

\subsection{Scene content}
	\label{ssec:scene_content}
	
	\begin{mdframed}[style=observation]
		\textbf{Scene content.} Edges in natural images are predominately horizontal and vertical. In popular image datasets, horizontal edges are more prevalent than vertical edges.
	\end{mdframed}

	\noindent\textbf{Background}\indent
	Most photographs show visual recordings of our world and therefore inherit the orientation characteristics of the environment. 
	Since the 1970s, researchers have measured the distribution of oriented contours in indoor, outdoor, and natural scene environments.
	They have found that contours near the cardinal axes, i.e., vertical and horizontal edges, are more prevalent than contours with oblique orientation angles~\cite{Switkes1978Frequency, Coppola1998OrientationDistribution}. 
	This holds especially for urban outdoor and indoor scenes, but also for natural scenes.
	Extensive analyses of horizontal vs.\ vertical differences support the finding that horizontal lines are more prevalent than vertical lines~\cite{Keil2000Oblique,Hansen2004HorizontalBias}.

	The prevalence of cardinal contours is also reflected in human perception. 
	Neurophysiologists have found that humans perceive horizontal and vertical lines more readily than oblique lines. 
	This is known as the \emph{oblique effect}~\cite{Appelle1972ObliqueEffect}.
	This bias can be linked to the physiology of the visual cortex, which has more cells processing horizontal and vertical than oblique information~\cite{Chapman1996VisualCortex}. 
	Other work discusses the interaction between the living environment and the development of visual perception. 
	The so-called carpentered world hypothesis suggests that people in urban environments develop stronger preferences for cardinal lines than people from rural areas~\cite{Segall1966CarpenteredWorld}.
	Largely independent of the living environment is the visual perception of human faces.
	A recent study presented filtered versions of human faces to subjects in fMRI brain scanners.
	It found that stimuli with horizontal frequencies activated the primary visual cortex more than vertical or oblique frequencies~\cite{Goffaux2016Neuropsychologia}. (Examples of the stimulus material are in Appendix~\ref{app:stimuli}.)
	The prevalence of cardinal contours is also present in the art.
	Analyses of 20th century paintings show that artists prefer horizontal and vertical over oblique lines, which may suggest that human observers find cardinal lines more aesthetically pleasing~\cite{Latto2002Paintings}. 
	In particular, paintings in portrait format contain more vertical lines while paintings in landscape format contain more horizontal lines~\cite{Latto2002Paintings}.
	
	\begin{figure}
		\includegraphics[width=\columnwidth]{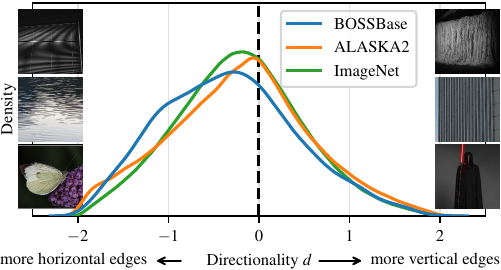}
		\caption{The relative distribution of horizontal and vertical scene content is skewed towards images with more horizontal edges. On the sides are example images from the three databases with very high absolute directionality scores.}
		\label{fig:dataset_directionality}
	\end{figure}

	\paragraph{Validation} We analyze the relative distribution of horizontal and vertical scene content in the BOSSBase, ALASKA2, and ImageNet databases using steerable pyramids. 
	Figure~\ref{fig:dataset_directionality} shows the distribution of directionality in the three databases. 
	The directionality score $d$ on the x-axis ranges between $-2$ and $+2$. 
	Images with negative $d$ contain more horizontal edges, images with positive $d$ contain more vertical edges. 
	While all databases contain diverse images, the distribution is skewed towards images with horizontal edges.
	These findings are consistent with related work~\cite{Hansen2004HorizontalBias}.
	The number of images with $\left| d \right| > 0.5$ ranges between $49\,\%$ and $57\,\%$.
	Through manual inspection, we observe that BOSSBase contains many seascapes with a clear horizon and horizontal edges.
\subsection{Sensor noise}
	The sensor consists of a rectangular array of photo diodes, which convert incoming photons to an electric charge. 
	The charges are collected, read from the sensor, amplified, and eventually converted to a digital signal~\cite{Fridrich2013Prnu}. 
	Imperfections in this process introduce noise into the resulting output image. 
	The exact sources of noise and its visual manifestation depend on the sensor design, which is usually proprietary. 
	While some sources of noise cause barely perceptible white noise, others lead to banding, which is more perceptible to the human eye.
	To give one example, we illustrate how the readout of CMOS sensors produces column patterns~\cite{ElGamal1998CmosFixedPatternNoise, Aguerrebere2012Acquisition}.
	
	\begin{mdframed}[style=observation]
		\textbf{Readout noise.} Differences between individual column amplifiers in CMOS sensors produce a column pattern.
	\end{mdframed}
	
	\noindent\textbf{Background}\indent
	In CMOS sensors, each individual cell converts the charge to a voltage. The voltages are then read line by line after shifting the charges through the column amplifier. Deviations in the column amplifiers introduce a column pattern. This column pattern can be best seen from bias frames.

	\begin{figure}
		\includegraphics{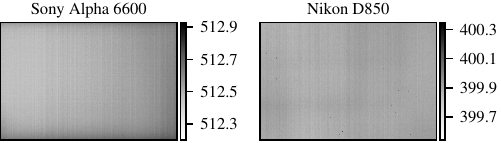}
		\caption{Physical differences between column amplifiers in CMOS sensors produce column artifacts.}
		\label{fig:cmos_column_noise}
	\end{figure}
	
	\paragraph{Validation}
	To illustrate the column pattern, we recorded 200 raw images with a Sony Alpha 6600 and a Nikon D850 camera in a dark room with the lens cap on. 
	Both cameras save 14 bits per pixel.
	The exposure time was set to the minimum of $1/4000\,s$ and $1/8000\,s$, respectively. 
	The short exposure time keeps the influence of dark current low and therefore allows us to measure the readout noise only.
	Figure~\ref{fig:cmos_column_noise} shows the average over the 200 frames. 
	The column pattern is clearly visible in the average frame, however, its amplitude is rather small. 
	The values fluctuate around $512$ and $400$, which are the two cameras' black levels.
	
\subsection{Raw-to-image conversion}
	\begin{mdframed}[style=observation]
		\textbf{Linear pattern.} Camera model-specific processing steps leave characteristic horizontal and vertical traces.
	\end{mdframed}
	
	\noindent\textbf{Background}\indent
	The raw sensor measurements undergo a sequence of processing steps to produce a final image. 
	They include color filter array (CFA) demosaicing, tone mapping, white balancing, denoising, sharpening, and often compression. 
	While the exact process in commercial cameras is proprietary and thus difficult to analyze, its effect can be observed by averaging many frames of the same camera model.
	
	The sensor together with the raw-to-image conversion introduce a so-called linear pattern, which can be obtained as a byproduct of the camera fingerprint estimation. 
	Camera fingerprints are estimated by averaging noise residuals from dozens of images, %
	followed by zero-meaning the rows and columns separately for each channel in the color filter array, and then applying a Wiener filter~\cite{Fridrich2013Prnu}.
	The difference between the average noise residual and the final fingerprint gives the linear pattern~\cite{Chen2007LinearPattern}. 
	It contains artifacts which are shared by cameras of the same model, unlike the fingerprint, which is intended to uniquely identify each sensor~\cite{Filler2008CameraModelIdentification}.
	
	\begin{figure}
		\includegraphics{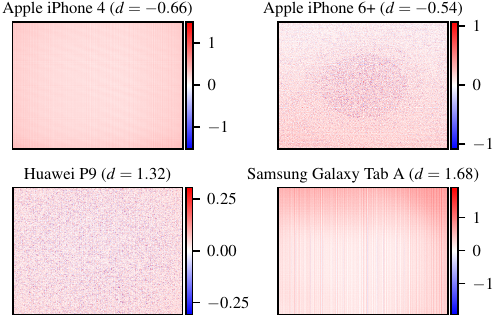}
		\caption{Spatial representation of linear patterns and directionality score $d$ of four cameras in the VISION database.}
		\label{fig:linear_patterns}
	\end{figure}
	
	\paragraph{Validation}
	The flat-field frames in the VISION database are particularly suited for studying linear patterns of different cameras~\cite{Shullani2017Vision}.
	Figure~\ref{fig:linear_patterns} shows four selected examples.
	Observe that the Apple iPhone~4 leaves strong horizontal banding artifacts.
	Apple's iPhone~6+ shows an oval shape in the center.
	The Huawei P9 shows irregular vertical structures, while the Samsung Galaxy Tab A leaves strong vertical banding artifacts.
	The directionality score $d$, calculated with steerable pyramids, confirms the visual impression.
	Although it is difficult to pinpoint the exact causes in these proprietary devices, these artifacts contribute to directional statistics in final images.
	We conjecture that the oval shape of the iPhone~6+ might be related to the correction of geometric lens distortion, another cause of directional artifacts which we discuss next.
	
	\begin{mdframed}[style=observation]
		\textbf{Geometric distortion correction.} Correcting geometric lens distortion introduces resampling artifacts, which are strongest at the longer ends of rectangular images.
	\end{mdframed}
	
	\noindent\textbf{Background}\indent
	An optical lens focuses incoming light onto the camera sensor.
	An ideal lens would map straight lines onto straight lines. 
	Most practical lenses are not ideal. 
	Their magnification changes with the distance from the optical center~\cite[Ch.~6.3]{Hecht2016Optics}. %
	As a result, straight lines appear curved. 
	We distinguish barrel from pincushion distortion, where the magnification decreases or increases with distance from the optical axis, respectively.
	While this geometric distortion is radial (i.e., non-directional), rectangular sensors capture more distortion at the longer ends of the frame. 
	This raises the question why most cameras have rectangular sensors. 
	For example,  DSLR full frame sensors typically measure $36 \times 24$\,mm. 
	This 3:2 aspect ratio has likely persisted from the \SI{35}{\milli\meter} film.
	
	A simple calculation also shows that rectangular sensors are more economical when photographers prefer rectangular aspect ratios.\footnote{Rectangular formats can be related to artistic rules, like the rule of thirds, and the golden ratio. In historical paintings, where artists were allegedly free to choose the aspect ratio, the majority of paintings are rectangular rather than square. \url{https://blog.wolfram.com/2015/11/18/aspect-ratios-in-art-what-is-better-than-being-golden-being-plastic-rooted-or-just-rational-investigating-aspect-ratios-of-old-vs-modern-paintings/} (accessed on 2 Feb, 2024).} 
	Lenses produce a spherical projection. 
	Since spherical sensors are difficult to manufacture, a square shape would maximize the area within a spherical projection. 
	However, cropping the square frame to the desired rectangular shape would imply losing resolution. 
	Directly fitting a 3:2 rectangle into the spherical lens projection gives $38$\,\% more area than fitting a square into the same size and cropping to the 3:2 format afterwards. %
	
	Most cameras accept geometric distortion on the optical side and correct for it using signal processing.
	This involves unwarping the image, a geometric transformation that requires interpolation~\cite[Ch.~4.2]{Farid2016PhotoForensics}.
	Interpolation, in turn, leaves statistical measurable correlations between adjacent pixels~\cite{Popescu2005Resampling}.
	
	\begin{figure}
		\begin{subfigure}{.45\columnwidth}%
			\centering
			{\fontsize{8}{10}\selectfont Sony E35 mm (\SI{35}{\milli\meter})}\\
			\includegraphics[width=\textwidth]{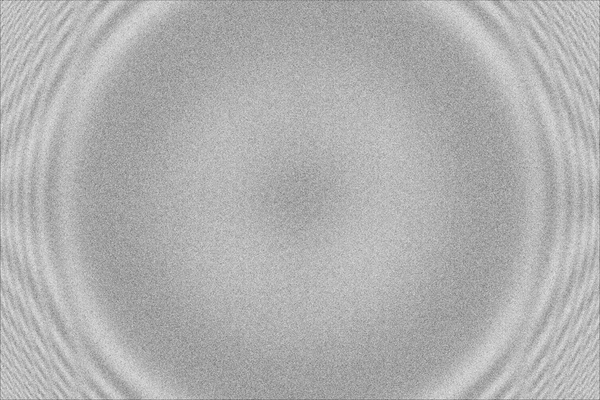}%
		\end{subfigure}%
		\hfill%
		\begin{subfigure}{0.45\columnwidth}%
			\centering
			{\fontsize{8}{10}\selectfont Nikkor 24-70 (\SI{38}{\milli\meter})}\\
			\includegraphics[width=\textwidth]{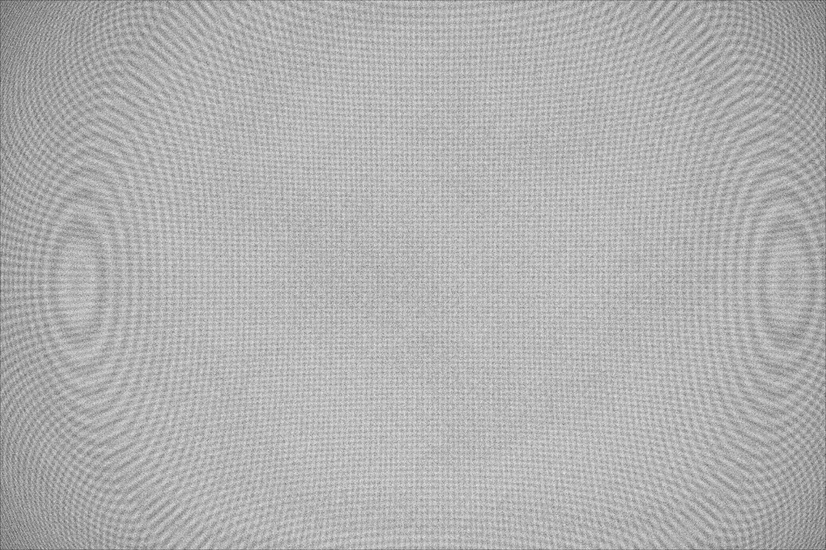}
		\end{subfigure}%
		\caption{Algorithmic correction of geometric lens distortion may introduce directionality as it treats the long ends of rectangular frames differently than the center area. 
		The figure illustrates this by showing the amount of local correlation for each pixel (brighter means stronger). 
		See text for details.}
		\label{fig:lens_correction_pmap}
	\end{figure}
	
	\paragraph{Validation}
	For illustration, we take two raw flat-field images with a Sony E 35\,mm prime lens and a Nikkor 24--70 zoom lens at 38\,mm, and correct for geometric distortion in Adobe Lightroom Classic 12.3.
	The corrected images are analyzed with a forensic resampling detector~\cite{Popescu2005Resampling}. 
	Figure~\ref{fig:lens_correction_pmap} shows the resulting $p$-maps.
	Brighter shades in $p$-maps mean that pixels can be better predicted from their local neighborhood.
	Known variations in local predictability can be exploited for more reliable forensics and steganalysis~\cite{Kodovsky2013ResizedCovers}.

\subsection{JPEG compression}
	\label{ssec:causes_jpeg}

	JPEG is still the most popular image format supporting lossy compression~\cite{Dornauer2023Formats}. 
	It achieves its storage efficiency by discarding visually unimportant information. 
	JPEG compression proceeds as follows~\cite{Pennebaker1992JPEGStandard}.  %
	RGB pixel values are converted to the luminance and chrominance color space. 
	The chrominance channels are optionally subsampled. 
	Every channel is split into non-overlapping $8\times8$ blocks, each of which is transformed to the frequency domain using the discrete cosine transform~(DCT). 
	The resulting DCT coefficients are quantized by division with subband-specific quantization factors and subsequent rounding.
	The 64 quantization factors form a quantization table~(QT), which can be different per channel, and is typically derived using the quality factor (QF) parameter in the (useful) range between 50 (low quality, small files) and 100 (all factors set to $1$, high quality, but still not lossless). 
	The quantized DCT coefficients are stored using lossless entropy encoding.
	JPEG decompression works in reverse order. 
	Information discarded after rounding is not recoverable.
	
	Although the standard does not introduce any directional artifacts, we find several sources of directionality in common parameter choices and in popular implementations.
	
	\begin{mdframed}[style=observation]
		\textbf{Chroma subsampling.}\;
		Horizontal chroma subsampling is more popular than vertical chroma subsampling. 
		Moreover, the popular libraries \turbo{} and \mozjpeg{} introduce horizontal high-frequency  artifacts.
	\end{mdframed}

	\noindent\textbf{Background}\indent
	Human perception is less sensitive to changes in color than to changes in brightness~\cite[Sec.~4.3.4, p.~221]{Reinhard2008ColorImaging}.
	JPEG compression exploits this property by optionally reducing the resolution of the chroma channels. 
	There are several common subsampling factors, usually denoted in J:a:b notation~\cite[Ch.~10]{Poynton2003Video}. %
	Here it is sufficient that 4:2:0 halves the chroma channel resolution in both directions, i.e., to $1/4$ of the pixels;
	horizontal subsampling (4:2:2) halves the horizontal resolution while keeping the vertical resolution; and
	vertical subsampling (4:4:0) halves the vertical resolution while keeping the horizontal resolution.
	
	Among the directional parameter choices, horizontal subsampling (4:2:2) is more common than vertical (4:4:0).
	This is probably due to the use of interlacing on cathode ray tube~(CRT) monitors, and in accordance with the ITU-R~BT.601 recommendation in favor of horizontal chroma subsampling~\cite{ITU2011StudioEncoding}. %
	Interlacing was adopted in the 1930s to double the frame rate without increasing the bandwidth requirements, thereby reducing temporal flickering from low frame rates~\cite[pp.\ 426--427]{Burns1998Television}.
	Interlacing splits the frame into two fields with alternating rows. 
	Only one field is updated at a time. 
	Decompressing a horizontally subsampled field is easy by replicating the previous chroma sample in a row. 
	By contrast, decompressing and displaying a vertically subsampled frame is more involved. 
	As CRT monitors update row by row, additional memory would be required to buffer a row of chroma samples.\footnote{Personal communication with Eric Bourguignat of (formerly) CCETT, a research center in Rennes, France, that contributed to the JPEG standard.} %
	These are all historical reasons. 
	Today's displays do not use interlacing, memory has become cheaper, and bandwidth has multiplied.
	
	\xmakefirstuc{\libjpeg}~\cite{libjpeg}, the open-source reference implementation, introduced a high-frequency horizontal artifact during chroma subsampling caused by integer rounding~\cite{Lorch2019ChromaWrinkles}. This artifact, called \emph{chroma wrinkle}, was found in version~6b, which has been in use for about a decade (1998--2009)~\cite{Benes2022KnowYourLib}. 
	From version~7 onwards, \libjpeg{} resolved this problem by changing the chroma subsampling implementation to DCT scaling. 
	Yet, the artifact persists until today in popular forks like \turbo{}~\cite{libjpeg-turbo}, which uses SIMD, and \mozjpeg{}~\cite{mozjpeg},
	which is optimized for web publishers~\cite{Hofer2023Progressive}. %
	
	\paragraph{Validation}
	\begin{table}
		\centering
		\caption{Number of cameras in forensic image datasets split by their use of chroma subsampling. Cameras in the ``mix'' column produce JPEGs with 4:2:0 and 4:2:2 subsampling.}  %
		\label{tab:chroma_subsampling_prevalence}
		\begin{tabular}{lcccc}
			\toprule
			\thead[l]{Database} & \thead{4:4:4} & \thead{4:2:0} & \thead{4:2:2} & \thead{mix} \\[-1ex]
			&&&\small (directional) \\
			\midrule
			Dresden & -- & 1 & 26 & -- \\
			VISION & -- & 32 & 2 & 1 \\
			Forchheim & -- & 21 & 5 & 1 \\
			\bottomrule
		\end{tabular}
	\end{table}
	
	We analyze the distribution of chroma sampling patterns in native camera images from forensic image databases. 
	The Dresden database~\cite{Gloe2010Dresden} contains JPEG images from DSLR and point-and-shoot cameras. The VISION~\cite{Shullani2017Vision} and Forchheim databases~\cite{Hadwiger2020Forchheim} contain smartphone photos. As shown in Tab.~\ref{tab:chroma_subsampling_prevalence}, 26 out of 27 cameras in the Dresden database use 4:2:2 subsampling. The only exception is the Kodak M1063, which uses 4:2:0 subsampling.
	In the VISION and Forchheim datasets, 4:2:0 subsampling is more prevalent. Still, some smartphones produce 4:2:2 subsampling.
	
	\xmakefirstuc{\libjpeg{}} and its derivatives include the command line tool \texttt{cjpeg} for compressing images. Interestingly, \mozjpeg{}'s \texttt{cjpeg} chooses the chroma subsampling based on the given quality factor. For $\text{QF} \geq 90$, \mozjpeg{} disables chroma subsampling (4:4:4), for $80 \leq \text{QF} < 90$, it subsamples horizontally (4:2:2), and for $\text{QF} < 80$, \mozjpeg{} defaults to 4:2:0.

	\begin{mdframed}[style=observation]
		\textbf{DCT implementation.}\;
		The integer DCT implementations in \libjpeg{} introduce directional rounding artifacts.
	\end{mdframed}

	\noindent\textbf{Background}\indent
	Since the 2-D DCT is linearly separable, it is commonly implemented as a row-wise 1-D DCT followed by a column-wise 1-D DCT~\cite{Agarwal2020RoundingArtifacts}.
	\xmakefirstuc{\libjpeg{}} supports three DCT methods.
	The default, called ISLOW, converts fractional constants to 13-bit integers and uses a fixed-point Loeffler--Ligtenberg--Moschytz~(LLM) algorithm, which saves multiplications by using a divide-and-conquer strategy~\cite{Loeffler1989Dct}.
	Fixed-point arithmetic requires descaling (bitwise shifts to the right) followed by rounding.\footnote{The scaling is explained in \textit{jfdctint.c} in \libjpeg{}.} 
	To maintain precision, this happens only after the second (column-wise) DCT.
	A property of the LLM algorithm is that the 1st and 5th element of the row vectors are not multiplied by fractional constants.
	These are upscaled by only 2 bits after the first pass, which is why the final descaling leaves stronger rounding artifacts in the 1st and 5th row, but no artifacts in the 1st and 5th elements.

	The faster but less accurate IFAST option uses the Arai--Agui--Nakajima~(AAN) DCT~\cite[p.~52]{Pennebaker1992JPEGStandard} algorithm with fixed-point arithmetic. 
	The implementation processes columns first and rows second.
	Compared to the ISLOW implementation, fractional constants are converted to only 8-bit integers and intermediate results are descaled immediately.
	
	The more accurate but slower IFLOAT DCT also uses the AAN DCT but with floats. We did not observe any noticeable rounding errors.
	DCT rounding artifacts are a cause of directionality, which is strongest at quality factor 100 as quantization attenuates these weak traces.
	
	\paragraph{Validation}
	\begin{figure}
		\begin{subfigure}[t]{\columnwidth}
			\centering
 			\includegraphics{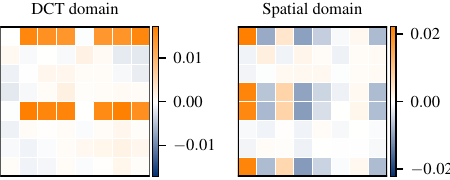}
			\caption{ISLOW DCT}
		\end{subfigure}\\
		\begin{subfigure}[t]{\columnwidth}
			\centering
			\includegraphics{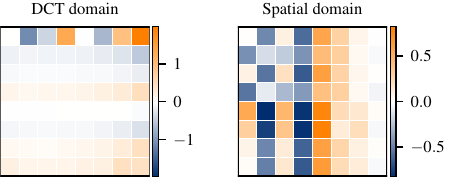}
			\caption{IFAST DCT}
		\end{subfigure}
		\caption{Practical DCT implementations in \libjpeg{} leave directional rounding artifacts. The top row is the default setting.}
		\label{fig:dct_artifacts}
	\end{figure}

	We draw \numprint{100000} random symmetric $8 \times 8$ blocks. 
	Each block is DCT-transformed using the ISLOW and IFAST DCT methods. 
	We measure the difference between the DCT block and a floating point version which does not have rounding artifacts.
	Figure~\ref{fig:dct_artifacts} shows the average difference of the ISLOW DCT~(top) and the IFAST DCT~(bottom) in the DCT and the spatial domain. 
	The DC difference is set to zero.
	Positive differences (orange) mean that the DCT implementation overestimates the corresponding coefficients.
	The ISLOW DCT overestimates rows 1 and 5 in the DCT domain.
	The differences are quite small.
	The IFAST DCT produces stronger artifacts, i.e., in spatial domain it underestimates columns 2 and 4 and overestimates column 5.
	
	\begin{mdframed}[style=observation]
		\textbf{Quantization tables.}\;
		Many quantization tables are asymmetric, including the ones \libjpeg{} uses by default.
	\end{mdframed}

	\noindent\textbf{Background}\indent
	The idea of subband-specific quantization is to attenuate those spatial frequencies that are barely perceptible. 
	The tables are stored in every JPEG file, therefore camera manufacturers and software developers can choose them.
	
	In Annex~K, the JPEG standard~\cite{ITU1993JpegRecommendation} provides example quantization tables~(QTs) for luminance and chrominance components. 
	They are the result of psychovisual thresholding experiments in which study participants were asked to indicate at what intensity the projection of each DCT basis function was discernible from a constant background~\cite{Lohscheller1984Visibility}. 
	Measurements from the original experiment are shown in Appendix~\ref{app:contrast_sensitivity}.
	Although the JPEG standard emphasizes that these two tables are only examples, their inclusion as defaults in \libjpeg{} has made them the de facto standard in many software implementations and in academic research. 
	These QTs are still in use today, although the luminance table has been criticized since the 1990s for its asymmetry and because certain higher frequencies are less quantized than lower frequencies~\cite{Klein1992HumanVision}.
	
	While the chrominance table is symmetric, the luminance table (shown in Fig.~\ref{fig:standard_qt}) has different values for horizontal and vertical frequencies. 
	For example, at 2/2 cycles per block, the horizontal quantization factor is $10$, whereas the vertical factor is $14$ (orange circles). 
	Conversely, at 5/2 cycles per block, the horizontal factor is $40$, whereas the vertical one is $24$ (blue).
	These seemingly minor differences affect JPEG steganography methods, such as J-UNIWARD~\cite{Holub2014Uniward} and UERD~\cite{Guo2015Uerd}, which embed more into coefficients with lower quantization factors.
	
	\paragraph{Validation}
	\begin{figure}
		\centering\small
		\begin{tikzpicture}[
			ampersand replacement=\&,
			highlight/.style={rounded corners,ultra thick, draw=black, opacity=0.3},
			]
			\matrix [matrix of math nodes] (m)
			{
				16 \& 11 \& 10 \& 16 \&  24 \&  40 \&  51 \&  61\\
				12 \& 12 \& 14 \& 19 \&  26 \&  58 \&  60 \&  55\\
				14 \& 13 \& 16 \& 24 \&  40 \&  57 \&  69 \&  56\\
				14 \& 17 \& 22 \& 29 \&  51 \&  87 \&  80 \&  62\\
				18 \& 22 \& 37 \& 56 \&  68 \& 109 \& 103 \&  77\\
				24 \& 35 \& 55 \& 64 \&  81 \& 104 \& 113 \&  92\\
				49 \& 64 \& 78 \& 87 \& 103 \& 121 \& 120 \& 101\\
				72 \& 92 \& 95 \& 98 \& 112 \& 100 \& 103 \&  99\\
			};

			\draw (m-3-1) node [draw,circle,thick,maincolor,inner sep=5pt] {~};
			\draw (m-1-3) node [draw,circle,thick,maincolor,inner sep=5pt] {~};

			\draw (m-6-1) node [draw,circle,thick,altcolor,inner sep=5pt] {~};
			\draw (m-1-6) node [draw,circle,thick,altcolor,inner sep=5pt] {~};
			
			\draw[->] ($(m-1-1.north west)+(0, 0.2)$) -- ($(m-1-8.north east)+(0, 0.2)$) node[midway, above, font=\small] {Horizontal frequencies};
			\draw[->] ($(m-1-1.north west)+(-0.2, 0)$) -- ($(m-8-1.south west)+(-0.2, 0)$) node[midway, above, font=\small, rotate=90] {Vertical frequencies};
		\end{tikzpicture}
		\caption{The luminance quantization table from Annex~K of the JPEG standard~\cite{ITU1993JpegRecommendation} has several asymmetric features.}
		\label{fig:standard_qt}
	\end{figure}
	
	\begin{table}
		\centering
		\caption{Prevalence of standard quantization tables~(QTs) and asymmetric luminance and chrominance QTs in native camera images from popular forensic datasets.}
		\begin{tabular}{lccc}
			\toprule
			\thead[l]{Dataset} & \thead{Standard\\QTs} & \thead{Asymmetric\\luma QT} & \thead{Asymmetric\\chroma QT} \\
			\midrule
			Dresden & 31\,\% & 96\,\% & 19\,\% \\
			VISION & 50\,\% & 68\,\% & 13\,\% \\
			Forchheim & 52\,\% & 99\,\% & 29\,\% \\
			\bottomrule
		\end{tabular}
		\label{tab:quantization_tables}
	\end{table}
	
	We analyze the JPEG images of three forensic image databases\footnote{We do not include the steganalysis databases because their original images are never compressed.} for the ratio of asymmetric QTs. 
	Table~\ref{tab:quantization_tables} shows that across all databases, the majority of luminance tables are asymmetric. 
	Conversely, the majority of chrominance tables are symmetric.
	Furthermore, we had access to \numprint{99677} user-supplied JPEG images~\cite{Park2018QuantizationTables}.
	Our own reanalysis revealed \numprint{1169} distinct luminance QTs including the standard tables ranging from QF~1 to QF~100. Out of these \numprint{1169} luminance QTs, 997 (85\,\%) are asymmetric.
	
	\xmakefirstuc{\libjpeg{}} calculates QTs by scaling the standard table~\cite{Yousfi2020CompressionQuality}.
	The only QF leading to a symmetric QT is 100, where all entries are 1. 
	\xmakefirstuc{\turbo{}} uses the same luminance QTs. %
	\xmakefirstuc{\mozjpeg{}} supports several QTs, including the standard tables. 
	However, the default has changed to a symmetric table,\footnote{\url{http://www.imagemagick.org/discourse-server/viewtopic.php?f=22&t=20333&p=98008\#p98008}, accessed on 2 Feb, 2024} which refines a previous proposal~\cite{Klein1992HumanVision}.
	On Ubuntu systems, the popular image libraries Pillow and OpenCV link against \libjpeg{} version 6b and \turbo{} version 8. 
	Hence, software built with these packages introduces chroma wrinkles and uses asymmetric QTs.
	
	\paragraph{Lossless rotation}\label{par:losslessrot}
	Most cameras produce JPEGs with a fixed orientation and store the sensor orientation with respect to the ground as EXIF metadata.
	This lets the viewer rotate the image after decompression~\cite{JpegClub2002ExifOrientation}.
	JPEG images can also be rotated without decompression, e.g., using the command line tool \texttt{jpegtran}. This also rotates the directional traces.

\section{Effect on steganalysis}
\label{sec:steganalysis}

	Having discussed the causes of directionality, we move on to implications for security applications.
	We start with steganalysis, the task of detecting steganography~\cite{Boehme2010}.
	We show for several steganography methods that state-of-the-art CNN-based detectors may fail if the stego images are rotated (Sec.~\ref{ssec:steganalysis_sensitivitiy}).
	We present and evaluate mitigations (Sec.~\ref{ssec:post_embedding_rotation}), and link the effect to specific causes of directionality (Sec.~\ref{ssec:overfitting}).

	\subsection{Sensitivity to rotated test images}
	\label{ssec:steganalysis_sensitivitiy}

		The ALASKA2 database is split into three disjoint subsets: 80\,\% of the images are used for training, 10\,\% for validation, and 10\,\% for testing.
		All images have a resolution of $512^2$ pixels and are compressed using \turbo{} 2.1.0 with quality factor $75$ and no chroma subsampling~(4:4:4).
		The corresponding QT is roughly the standard QT shown in Fig.~\ref{fig:standard_qt} divided by two.
		Steganography is simulated using the three state-of-the-art JPEG steganography methods nsF5~\cite{Fridrich2007nsF5}, UERD~\cite{Guo2015Uerd}, and J-UNIWARD~\cite{Holub2014Uniward} with embedding rate 0.4~bits per non-zero AC coefficient~(bpnzAC) into the luminance channel.
		This setup and rate corresponds to a commonly chosen evaluation setting in contemporary papers~\cite{Yousfi2020ImageNetPretrained, Yousfi2021SurgicalModifications, Cogranne2020JMiPOD}.
		
		We use the AdamW optimizer to train an EfficientNet-B0 with ImageNet-pretrained weights~\cite{Butora2021Pretraining}, batch size $32$,  dropout rate $0.25$, and learning rate $10^{-4}$.
		The training images in the \emph{no-rot} setup were randomly flipped on the horizontal and vertical axis for data augmentation, but not rotated by 90 or 270 degrees in order to preserve the directionality.
		The top section of Tab.~\ref{tab:rotation_augmentation} (\emph{no-rot}, QF 75) reports the accuracy on the original test images (org.) and after rotating the test images by 90 degrees (rot.).
		Comparing the three steganography methods on the original orientation, nsF5 is least secure and J-UNIWARD is most difficult to detect. 
		This confirms related studies~\cite{Boroumand2019SRNet, Butora2021Pretraining, Itzhaki2021Augmentation}.
		Importantly, if the test images are rotated, the performance of all three detectors drops between $2.9$ and $9.8$\,\%-pts.
		This means that steganographers can gain an advantage by transmitting rotated images.

\begin{table}
			\centering
			\caption{Sensitivity of steganalysis to directionality (top section). 
			The first two columns report the accuracy for quality factor 75, the last two columns for an artificially asymmetric QT blending quality factors 60 and 80.
			}
			\begin{tabular}{l@{~~~}l@{~~}cccc}
				\toprule
				&&\multicolumn{2}{c}{QF 75} & \multicolumn{2}{c}{QF 80\textbackslash60} \\
				&&\multicolumn{2}{c}{\small (Section~\ref{ssec:steganalysis_sensitivitiy})} & \multicolumn{2}{c}{\small (Section~\ref{sssec:artificial})} \\
				\cmidrule{3-4}
				\cmidrule(l){5-6}
				\thead[l]{Setup} & \thead[l]{Stego.} & \thead{org.} & \thead{rot.} & \thead{org.} & \thead{rot.}\\
				\midrule
				\multirow{3}{*}{\emph{no-rot}}
				& nsF5 & 0.977 & 0.879 & 0.976 & 0.665 \\
				& UERD & 0.922 & 0.893 & 0.927 & 0.557 \\
				& J-UNIWARD & 0.884 & 0.801 & 0.893 & 0.492 \\
				\midrule
				\multicolumn{6}{c}{\emph{Effect of rotation augmentation}}\\
				&&\multicolumn{2}{c}{\small (Section~\ref{sssec:standard})}\\
				\cmidrule{3-4}								
				\multirow{3}{*}{\emph{aug-rot}}
				& nsF5 & 0.978 & 0.976 & 0.977 & 0.975 \\
				& UERD & 0.926 & 0.925 & 0.914 & 0.913 \\
				& J-UNIWARD & 0.897 & 0.897 & 0.893 & 0.873 \\
				\cmidrule{2-6}
				\multirow{3}{*}{\emph{base-rot}}
				& nsF5 & 0.986 & 0.864 & 0.984 & 0.671\\
				& UERD & 0.927 & 0.876 & 0.938 & 0.543 \\
				& J-UNIWARD & 0.911 & 0.780 & 0.908 & 0.505 \\
				\bottomrule
			\end{tabular}
			\label{tab:rotation_augmentation}
		\end{table}

	\subsection{Mitigation with rotation augmentation}
		\label{ssec:post_embedding_rotation}
		Data augmentation attempts to reduce overfitting by training on multiple variants of the existing samples.
		A popular augmentation practice is to rotate the decompressed training images by multiples of 90 degrees~\cite{Ye2017Hierarchical, Boroumand2019SRNet, Itzhaki2021Augmentation, Zhang2022AutoAugmentation}.
		Since image statistics are directional, rotating the test images can prevent a CNN from exploiting directional embedding traces.
		To validate this, we add two experimental setups.
		
		\subsubsection{Standard quantization table}
		\label{sssec:standard}

			The second setup, denoted \emph{aug-rot}, is identical to the first, but the images are randomly rotated by multiples of 90 degrees in the spatial domain during training, as is typical for data augmentation.
			Hence, the detector must generalize to both the original and the transposed QT.

			In the third setup, \emph{base-rot}, we create three additional copies of the database, where the cover images are rotated by 90, 180, and 270 degrees prior to compression.
			The stego images are created from the rotated covers.
			This dataset has four times the number of cover and stego images.
			Table~\ref{tab:rotation_augmentation} reports the test accuracy for all setups and embedding methods.
			Both rotation setups slightly improve the detection accuracy over the \emph{no-rot} setup, presumably because the CNNs are exposed to a greater variety of images.
			In the \emph{aug-rot} setup, the detectors generalize to rotated images, because they have seen both orientations during training.
			This mitigates the weakness and removes the steganographer's advantage from sending rotated images.
			
			The highest accuracy on the original test images is achieved by training with the \emph{base-rot} setup. The gain in accuracy compared to the \emph{aug-rot} training ranges between $0.1$ and $1.4$\,\%-pts.
			Although our gains are small, they are within the range of improvements that have been reported in recent publications.
			For example, tweaking the EfficientNet-B0 architecture by removing pooling promises gains below  $1$\,\%-pt., but at the cost of tripling FLOPs and memory~\cite{Yousfi2021SurgicalModifications}.
			Unsurprisingly, the detectors in the \emph{base-rot} setup experience the largest drop in accuracy when tested with rotated images.
			This is probably because they have learned directional traces in the training data that have been introduced by JPEG compression.
		
		\subsubsection{Artificial QT with amplified asymmetry}
		\label{sssec:artificial}

			Although the standard QT contains many asymmetric features, the table as a whole does not exhibit a dominant direction.
			Stronger quantization in one horizontal coefficient might be compensated by another vertical coefficient.
			In order to link the drop in performance to directionality caused by JPEG compression, and to rule out alternative explanations, we repeat the experiments with a QT that has artificially strong directionality.
			Our artificial QT blends the QTs for QF~60 in the upper and QF~80 in the lower triangle, and their average on the diagonal.
			We chose this combination because it is close to QF~75.
			Note that the performance cannot be compared directly to the results for QF~75, because the difficulty of the steganalysis task significantly depends on the QT.
			Everything else is unchanged.

			The two rightmost columns in Tab.~\ref{tab:rotation_augmentation} (QF 80\textbackslash60) report the test accuracies for this experiment.
			As before, in the \emph{no-rot} and \emph{base-rot} setups, the CNNs do not generalize to rotated images.
			However, the performance drops are more pronounced. 
			The accuracy on UERD and J-UNIWARD gets close to random guessing.
			We interpret this as evidence that the directionality picked up by the CNN is to a large extent introduced during JPEG compression, and specifically by the directionality in the QT.
			Again, the \emph{base-rot} CNNs show improvements over the \emph{aug-rot} CNNs between $0.7$ and $2.4$\,\%-pts.

	\subsection{Overfitting to dataset directionality}
	\label{ssec:overfitting}

		\begin{figure}
			\includegraphics{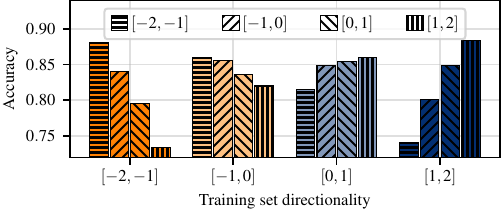} %
			\caption{
			Effect of dataset directionality on steganalysis.
			Each set of bars refers to one detector trained on artificially biased data with directionality scores in the reported interval.
			Hatched bars report the resulting accuracy on test data with varying directionality.
			Observe the overfit on the outer bins.
			}
			\label{fig:overfitting_directionality_alaska}
		\end{figure}
	
		Popular image databases, including the steganalysis databases BOSSBase and ALASKA2, are dominated by images with more horizontal than vertical edges (cf.~Fig.~\ref{fig:dataset_directionality}).
		When training on such biased databases, a learning-based steganalysis detector might overfit to the prevalent direction.
		We demonstrate this first on an artificial dataset and then on BOSSBase.
		
		\paragraph{Artificial dataset}
		For the first experiment, we chose the ALASKA2 database because it is larger than BOSSBase.
		The ALASKA2 database is split into 50\,\% for training and 50\,\% for testing. 
		Both subsets are split by directionality into four bins with the directionality scores $[-2, -1]$, $[-1, 0]$, $[0, 1]$, and $[1, 2]$. %
		From each bin, we randomly select the maximum number of images such that all bins have the same size (2484 training and 2592 test images).
		We then train a steganography detector with images from a single bin, and evaluate its test accuracy on all bins separately.
		To isolate the effect of the directionality of the scene content, we suppress the directional bias from JPEG compression by using a symmetrized variant of the QF-75 QT and the floating-point DCT method. 
		Like before, we embed steganography into the luminance channel using J-UNIWARD with 0.4 bpnzAC.
		Since we have too few images for training a CNN-based detector, we use feature-based steganalysis.
		We extract the popular Gabor filter residual~(GFR)~\cite{Song2015Gabor} features from the luminance channel and train an ensemble classifier with $\approx 100$ Fisher linear discriminants as base learners, as suggested in the literature~\cite{Kodovsky2012Ensemble}.

		Figure~\ref{fig:overfitting_directionality_alaska} shows the results.
		Each set of bars shows one ensemble classifier trained on one bin.
		The individual bars represent the accuracy on the individual test bins.
		When training with an extremely directional training set, i.e., $[-2, -1]$ and $[1, 2]$, the classifier achieves the highest accuracy on images with the same directionality.
		The accuracy drops when the test images contain a less pronounced or opposite directionality.
		When training with a moderately directional training set, i.e., $[-1, 0]$ and $[0, 1]$, the classifier shows higher accuracy for images of the same directionality than for the opposite direction.
		This experiment on an artificial subset of ALASKA2 demonstrates that even featured-based steganalysis is sensitive to a directionality bias in the training data.
		
		\paragraph{BOSSBase}
		Typical datasets are more diverse in their directionality than the one in the previous experiment.
		Now we demonstrate that the directionality bias in BOSSBase may impair into a steganography detector's generalizability.
		The BOSSBase images are split into 50\,\% for training and 50\,\% for testing.
		All images have a resolution of $512^2$ pixels and are compressed using \turbo{} 2.1.0 with the symmetric QF-75 QT and the floating-point DCT method.
		Steganography is simulated using nsF5~\cite{Fridrich2007nsF5}, UERD~\cite{Guo2015Uerd}, and J-UNIWARD~\cite{Holub2014Uniward} with an embedding rate of 0.4~bpnzAC.
		In this experiment, we use two popular steganalysis feature descriptors: the phase-aware spatial rich models~(PHARM)~\cite{Holub2015Pharm} and the Gabor filter residuals~(GFR)~\cite{Song2015Gabor}.
		Both feature sets do not merge horizontal and vertical orientations and can thus capture directional traces.
		We use the same ensemble classifier as before~\cite{Kodovsky2012Ensemble}.
		
		\begin{table}
			\caption{
			Effect of the directionality bias in the BOSSBase steganography benchmark dataset on detection accuracy.
			}
			\centering
			\begin{tabular}{llccc}
				\toprule
				\thead[l]{Steganography\\method} & \thead[l]{Steganalysis\\features} & \thead{Test\\all $d$} & \thead{Test\\ $d < 0$} & \thead{Test\\$d > 0$} \\
				\cmidrule{3-5}
				{\small \# images} && {\small 5000} & {\small 3417} & {\small 1583} \\
				\midrule
				\multirow{2}{*}{nsF5}
				& GFR & 0.993 & 0.993 & 0.991 \\
				& PHARM & 0.987 & 0.989 & 0.980 \\
				\midrule
				\multirow{2}{*}{UERD}
				& GFR & 0.894 & 0.900 & 0.879 \\
				& PHARM & 0.860 & 0.868 & 0.841 \\
				\midrule
				\multirow{2}{*}{J-UNIWARD}
				& GFR & 0.895 & 0.899 & 0.888\\
				& PHARM & 0.870 & 0.878 & 0.854\\
				\bottomrule
			\end{tabular}
			\label{tab:overfitting_dataset_directionality}
		\end{table}
		
		The left column in Tab.~\ref{tab:overfitting_dataset_directionality} shows the test accuracy over all test images. %
		Additionally, we break down the test set by the directionality score $d$. %
		The detector consistently achieves higher accuracy on test images with negative $d$.
		Although the difference is small, this result on a database that has served as a benchmark for many publications~\cite{Fridrich2012RichModels, Kodovsky2012Ensemble, Holub2014Uniward, Guo2015Uerd, Holub2015Pharm, Song2015Gabor, Ye2017Hierarchical, Boroumand2019SRNet} confirms that the detector works better on images that follow the database bias.
		We repeated this experiment on the original ALASKA2 database, but did not observe any significant differences in accuracy.
		This relates to the observation in Fig.~\ref{fig:dataset_directionality}, where we found that the directionality distribution in ALASKA2 is less skewed than in BOSSBase.
		ALASKA2 seems to be a better benchmark than BOSSBase in this regard.

\section{Effect on forensic source identification}
\label{sec:identification}

	Linking a photo to a camera make or model is a common forensic task relevant to law enforcement~\cite{Kirchner2015CameraModel}.
	This is possible because cameras differ in the technical implementation of the acquisition pipeline (cf.~Fig.~\ref{fig:acqusition_pipeline}).
	While recent advances in camera model identification propose CNNs~\cite{Bondi2017CamIdCnn, Rafi2020RemNet, Hadwiger2020Forchheim}, feature-based classifiers~\cite{Marra2015ResidualBasedLocalFeatures, Mendes2019OpenSetCamId} are still used, presumably because they are established, explainable (if necessary in court), and require fewer training images than CNNs.
	In this section, we show how considering directionality can improve the classification accuracy of feature-based methods.
	The results for the CNNs used in Fig.~\ref{fig:introduction} are in Appendix~\ref{app:cnn_based_camera_model_identification}.
	
	A popular feature descriptor for capturing camera model artifacts is the \emph{subtractive pixel adjacency model~(SPAM)}~\cite{Marra2015ResidualBasedLocalFeatures, Mendes2019OpenSetCamId}.
	SPAM features compute noise residuals and count co-occurrences between adjacent pixels. 
	Both happens in horizontal and vertical direction. 
	To decrease the feature dimensionality, certain horizontal and vertical co-occurrences are averaged~\cite{Pevny2010Spam}.
	This removes directional information.
	We show that feature-based camera model identification benefits from keeping directional image statistics separate.
	
	\begin{figure}
		\includegraphics{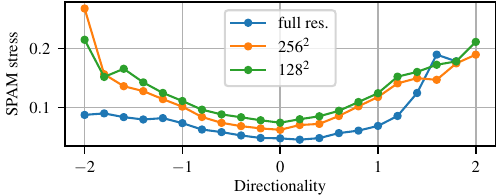}
		\caption{
			Relation between directionality and ``stress'' in the original SPAM feature space.
			Higher values indicate more information loss due to symmetrization.
			Our proposed variant removes this stress at the expense of greater dimensionality.
		}
		\label{fig:spam_stress}
	\end{figure}

	Similar to previous work~\cite{Marra2015ResidualBasedLocalFeatures}, we randomly select 10 camera models with at least two devices from the Dresden database~\cite{Gloe2010Dresden}. 
	To ensure that the classifier learns model-specific rather than device-specific traces, images from only one device are used for training. 
	The other devices are used for testing.
	To properly account for potential lossless rotation (cf.~Sec.~\ref{par:losslessrot}), we verify that 17 (out of 18) camera models store the images with a canonical orientation.
	One camera model rotated the images prior to compression.
	We also use its images as stored, because this preserves the compression direction.
	We randomly select the same number of images from each camera model, such that training and test sets are class-balanced. 
	To avoid leakage from the scene content, $80\,\%$ of the scenes are exclusively used for training and the remaining $20\,\%$ are used for testing.
	
	We extract the SPAM features from each image.
	The original feature descriptor with directional averaging has 338 dimensions. 
	Removing the directional averaging yields 676 dimensions.
	The feature extraction is repeated for the full-resolution images and for center crops with a size of $256^2$ and $128^2$.
	Normalized features are used to train a linear support vector machine~(SVM), as suggested in previous work~\cite{Marra2015ResidualBasedLocalFeatures}.
	The regularization hyper-parameter of the SVM is set via a grid search with five-fold cross validation.
	Table~\ref{tab:forensic_camera_model_identification} shows the classification accuracy for this 10-class task.
	The results are averaged over five training runs with randomly selected camera models.
	At the full image resolution, both feature descriptors attain the same near-perfect accuracy.
	For the more challenging task on smaller image patches, the classifier benefits from the directional information in our extended feature descriptor.
	At size $128^2$, our non-symmetrized variant outperforms the original by $3$\,\%-pts.
	We observed similar results when using only the straight submodel or the diagonal submodel and with different classifiers.
	
	An advantage of feature-based methods is that we can measure the information lost by symmetrization.
	Figure~\ref{fig:spam_stress} plots the L2 distance between the submodels of different directionality that are averaged in the original SPAM features as a function of the directionality score.
	We call this distance ``stress.''
	The two submodels are most similar for directionality scores close to zero. 
	Stress increases with image directionality.
	Taken together, this provides the link between the directionality score, symmetrization in feature space, and accuracy.

	\begin{table}
		\caption{Effect of directionality on forensic camera model identification.
		The original SPAM features average horizontal and vertical information (symmetrization). 
		Skipping this step improves the classification accuracy for small images.}
		\centering
		\begin{tabular}{lcc}
			\toprule
			\thead[l]{Resolution} & \thead{SPAM~\cite{Pevny2010Spam}} & \thead{Proposed variant} \\
			& {\small{(symmetrized)}} & {\small{(not symmetrized)}} \\
			\midrule
			Full resolution & $0.9972$ & $0.9972$\\
			$256^2$	& $0.9722$ & $0.9728$ \\
			$128^2$	& $0.9041$ & $0.9342$ \\
			\bottomrule
		\end{tabular}
		\label{tab:forensic_camera_model_identification}
	\end{table}

\section{Effect on synthetic image detection}
\label{sec:synthetic}

	\label{sec:synthetic_image_detection}
	GAN and diffusion-based synthetic image generators are known to introduce weak but detectable periodic artifacts~\cite{Corvi2023SyntheticImageAnalysis}.
	They are explained by the repeated upsampling during image synthesis and therefore appear largely symmetrical~\cite{Frank2020Frequency}.
	Little attention has been paid to asymmetries in these artifacts.
	We show that synthetic images do contain directional traces, which CNN-based ``fake-or-real'' detectors tend to exploit.

	We train CNN detectors for two synthetic image generators: DALL·E Mini~\cite{Dayma2021DalleMini} and Stable Diffusion XL~(SDXL)~\cite{Podell2023StableDiffusionXL}.
	For the real images, we use ALASKA2 and rotate each image by a random multiple of 90 degrees to remove directional bias.
	To create fake images, we pass the real images through the autoencoders of the two synthetic image generators.
	Compared to generating images from text prompts, this approach allows us to create real--fake pairs with matching scene content. 
	This eliminates the possibility of spurious results due to shortcuts caused by a scene bias~\cite{Geirhos2020Shortuct}.
	As detailed in Appendix~\ref{app:autoencoder_vs_prompt}, this approach leaves the same synthesis artifacts as the typical text-to-image pipepline.
	$80\%$ of the images are used for training, $10\%$ for validation, and $10\%$ for testing. 
	We ensure that both real and fake images of the same scene are in the same split.
	We train an EfficientNet-B0 as binary classifier with cross-entropy loss, ImageNet pretraining, learning rate $10^{-4}$, batch size 32, and dropout rate $0.25$.
	
	\begin{table}
		\caption{CNN-based synthetic image detectors exploit directional artifacts introduced by the autoencoders of GANs.}
		\centering
		\begin{tabular}{l@{~}c@{~~}ccc}
			\toprule
			\thead[l]{Autoencoder} & \thead{Orientation} & \thead{Acc.} & \thead{TNR} & \thead{TPR} \\
			\midrule
			\multirow{2}{*}{DALL·E Mini} & original & $1.000$ & $1.000$ & $0.999$ \\
			& rotated & $0.989$ & $0.999$ & $0.979$\\
			\midrule
			\multirow{2}{*}{Stable Diffusion XL} & original & $1.000$ & $1.000$ & $1.000$ \\
			& rotated & $0.674$ & $1.000$ & $0.349$ \\
			\bottomrule
		\end{tabular}
		\label{tab:synthetic_image_detection_test_rotation}
	\end{table}

	Table~\ref{tab:synthetic_image_detection_test_rotation} reports that both detectors achieve an accuracy of $1.0$.
	The accuracy drops when the test images are rotated by 90 degrees.
	As can be seen from the true negative rate~(TNR) and true positive rate~(TPR), the performance loss is largely due to fake (positive) images being misclassified as real (negative).
	For the DALL·E Mini autoencoder, the TPR drops from $0.999$ to $0.979$, which means that $2.1\,\%$ of fake images are classified as real.
	For the SDXL autoencoder, the TPR even drops to $0.349$, which means that almost two out of three fake images are classified as real.
	Rotated real images are correctly classified, i.e., the TNR remains near-perfect after rotation.
	
	\begin{figure}
		\includegraphics{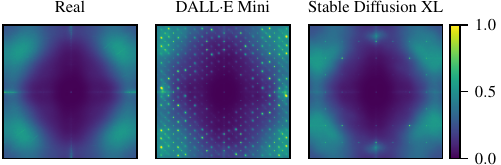}
		\caption{Power spectra of noise residuals from real images, the DALL·E Mini autoencoder, and the Stable Diffusion XL autoencoder. 
		The last stages of generative models leave asymmetric artifacts that do not appear in real images.
		}
		\label{fig:power_spectrum_autocorrelation}
	\end{figure}
	
	Directional artifacts can be observed in the power spectrum after averaging the noise residuals from \numprint{1000} randomly selected images~\cite{Corvi2023SyntheticImageAnalysis}. Figure~\ref{fig:power_spectrum_autocorrelation} compares the randomly rotated ALASKA2 (left) to our DALL·E Mini (center) and SDXL (right) images.
	In ALASKA2, no directional pattern can be observed.
	Conversely, DALL·E Mini shows asymmetric dot patterns.
	SDXL introduces a high-frequency artifact on the vertical axis, which is not present on the horizontal axis.

	Our experiments in this section demonstrate that synthetic images contain directional statistics.
	They are related to the autoencoder of synthetic image generators, but their root cause is yet unknown.
	As many recent detectors for synthetic images do not rotate their training images~\cite{Wang2020EasyToSpot, Cozzolino2021Universal, Gragnaniello2021CriticalAnalysis, Shiohara2022BlendedImages, Wang2023DiffusionDetection}, these detectors may be less reliable on rotated test images.
	An attacker can generate an image of a rotated scene and manually rotate it back to reduce the chances of being caught. %

\section{Related work}
\label{sec:related}

	Previous work in the security field has touched on the directionality of images, but the results are scattered.
	Westfeld~\cite{Westfeld2005Curves} reports improved steganalysis by scanning images along a space-filling Hilbert curve. 
	While it breaks up directionality, the method is justified by increased locality.
	Franz et al.~\cite{Franz2000CoverStegoAttacks, Franz2002Diss, Franz2008PixelDependencies} develop statistical models of flatbed scanner images.
	They observe directional differences between successive scans and attribute it to the variability of line sensor elements and the mechanical tolerances.
	None of this applies to the area sensors in today's smartphones and digital cameras.
	A common forensic task on these sensors is to determine the configuration of the color filter array. 
	The fact that this pattern depends on the camera orientation has been studied~\cite{Kirchner2010Cfa}.
	Recent work observed directional artifacts in DALL·E 2, but did not investigate further~\cite{Bammey2024Synthbuster}.
	
	Popular feature descriptors in steganalysis~\cite{Pevny2010Spam,Fridrich2012RichModels,Kodovsky2012JRM} and image forensics~\cite{Chen2015DemosaicingFeatures} use directional high-pass filters to extract noise residuals or count co-occurrences in horizontal and vertical directions.
	They exploit sign and mirror symmetries to obtain a compact feature vector.
	Several early works made the simplifying assumption that image statistics are non-directional;
	examples include the subtractive pixel adjacency model~(SPAM) (``the effect of portrait/landscape orientation is negligible''~\cite[p.~3]{Pevny2010Spam}), the spatial rich model (``the statistics of natural images do not change after rotating the image by 90 degrees''~\cite[p.~5]{Fridrich2012RichModels}), and the JPEG rich model (``assuming the statistics of natural images do not change after mirroring about the main diagonal''~\cite[p.~2]{Kodovsky2012JRM}).
	Such assumptions can be traced back to early quantitative detectors of LSB replacement steganography (``What holds when scanning an image horizontally ought to hold, in general, when scanning vertically.''~\cite[p.~13]{Ker2004Pair2D}).
	While a lower feature dimensionality reduces the risk of overfitting and multicolinearity~\cite{Yedroudj2018Augmentation}, %
	it ignores the natural directional properties of images. %
	Two more recent steganalysis feature descriptors, the phase-aware rich models~(PHARM)~\cite{Holub2015Pharm} and the Gabor filter residuals~(GFR)~\cite{Song2015Gabor}, do not average over orientations, but without giving a reason or supporting evidence.
	Some applications may ignore directionality when the orientation is known a priori or when the error rate is low enough to analyze the image in all four orientations, e.g., in PRNU matching.

	The practices regarding rotation augmentation are inconsistent among subfields.
	In steganalysis, rotation augmentation is common~\cite{Ye2017Hierarchical, Boroumand2019SRNet, Itzhaki2021Augmentation}.
	In camera model identification and synthetic image detection, the majority of related work avoids rotation, with two exceptions~\cite{Hadwiger2020Forchheim, Mandelli2022Orthogonal}. 
	We did not find reflections on these choices, let alone comparisons.
	Rotation augmentation has been adopted from computer vision~\cite{Lecun1995learning}, where images are often downscaled and rotated by only a few degrees. 
	However, these operations involve resampling which suppresses the signal of interest for security applications~\cite{Itzhaki2021Augmentation}.

	Overall, directionality of image statistics seems tangential to all of these works.
	While the security implications of certain image processing operations have received some attention recently (e.g., adversarial scaling~\cite{Quiring2020AdversarialScaling}), a comprehensive treatment of directionality has been lacking.

\section{Discussion}
\label{sec:discussion}

	Directionality matters.
	This is the two-word summary of our multi-domain, multi-method study on the causes and effects of directionality in digital images in security applications.

	This research has led to a number of new insights, primarily into the effects but also the causes of directionality.
	We are the first to report directional bias in popular benchmark image databases. 
	We also point out that DCT implementations in popular JPEG libraries leave directional traces. %
	For color images, directional chroma subsampling and asymmetric quantization tables add more directional traces.
	Across the three security applications studied, detectors trained on one image orientation do not naturally generalize to rotated images.
	In steganalysis, we found that the main cause for this is asymmetric quantization tables, followed by directional scene bias in the dataset.
	CNN-based fake-or-real detectors are susceptible to pick up directional traces left by the autoencoders in the last stage of generative models.
	Several feature descriptors used in steganalysis and image forensics average horizontal and vertical statistics. 
	We found that a forensic camera model classifier benefits from keeping the directions separate, especially for small images.
	
	Rotation augmentation requires more thought than is devoted in the extant literature.
	Our results lead to the following recommendations.
	If generalization to rotated images is desired, use rotation augmentation.
	To maximize performance on a directional dataset, avoid rotation augmentation.
	On small datasets, rotation augmentation can reduce overfitting but directionality-preserving transformations are preferable.
	
	A qualitative observation is that many causes of directionality relate to legacy conventions and practices.
	While old standards ``never die,'' and are notoriously hard to replace, contemporary implementations of standards could depart from outdated defaults and instead use settings that do not leave new or amplify existing directional traces. 

	\vspace{-1ex}
	\paragraph{Limitations}
	While we claim to have captured the dominant causes of directionality in raw-to-image conversion, our stock-taking is probably incomplete.
	A careful study of the literature reveals more idiosyncrasies of specific devices that may cause directionality.
	For example, the meanwhile historical Fujifilm J50 reportedly shifts the PRNU by seven pixels in horizontal direction depending on the exposure time~\cite{Gloe2012PrnuArtifacts}.
	
	Although the experimental setups in Sections~\ref{sec:steganalysis}, \ref{sec:identification}, and \ref{sec:synthetic} closely replicate selected related work published in prominent venues, these scenarios only evaluate a single dataset, compression setting, and detector architecture.
	The effect of directionality may be larger or nonexistent in different scenarios.
	Additionally, we trained and evaluated our CNNs with JPEG images that were \emph{de}compressed with default settings, i.e., the ISLOW inverse DCT.
	Switching to the float DCT would slow down the training and the numerical differences are tiny.
	Real-world applications built with the popular TensorFlow framework might deviate much more since \texttt{tf.io.decode\_jpeg} defaults to the IFAST DCT method.\footnote{See \url{https://github.com/tensorflow/tensorflow/blob/907415fea3e5e8ae2afee09c42f3cc1152fe63fd/tensorflow/core/kernels/image/decode_image_op.cc\#L127}, accessed on 31 Jan, 2024}
	The CNN detectors in Sec.~\ref{sec:synthetic_image_detection} are tailored to \emph{un}compressed synthetic images. 
	In practical applications, suspect images are often post-processed, scaled, and compressed. 
	We intentionally chose uncompressed images to avoid potential inferences from directional effects of subsequent processing.
	Further experiments with more diverse setups are needed to quantify which factors interact with directionality.

	\vspace{-1ex}
	\paragraph{Possible extensions}
	Directionality relates to rotation by 90 degrees.
	Is there a similar paper to be written on horizontal and vertical flipping?
	Clearly, natural images may contain biases, e.g., a bright sky at the top and darker areas at the bottom. 
	Some processes intentionally flip images, e.g., webcams make users feel like they are looking into a mirror.
	To show that directionality is still special, we retrain all CNNs from Fig.~\ref{fig:introduction} without flipping augmentation, and evaluate their accuracy with flipped images. 
	In all cases, the performance on flipped images matches the performance on the original images. 
	This suggests that flipping matters less than rotation.

	\vspace{-1ex}
	\paragraph{Reproducibility} Our code for measuring directionality is available at \href{https://github.com/uibk-uncover/directionality}{https://github.com/uibk-uncover/directionality}.

\section{Concluding remark}
\label{sec:conclusion}
	We close with the conjecture that gravitation is the root cause of directionality in digital images. 
	Human preference for landscape orientation has impacted scene composition as well as technology. 
	This is likely due to the horizontal arrangement of mammalian eyes, which itself might have evolved to make our prehistoric ancestors more secure against predators that move on the surface~\cite{Banks2015Pupils}.
\subsection*{Acknowledgements}
This work is funded by the European Union's Horizon 2020 research and innovation programme under grant agreement No.~101021687 (UNCOVER).
The numerical results were calculated on the University of Innsbruck's LEO5 cluster and the Vienna Scientific Cluster~(VSC).
We thank our colleagues Nora Hofer, Martin Bene{\v s}, and Verena Lachner for helpful comments on the study, and Alain L\'{e}ger, Eric Bourguignat, and Herbert Lohscheller for discussions on the JPEG history.
\bibliographystyle{plain}
{\small\bibliography{\jobname}}

\appendix

\section{Data}
	\label{app:data}

	The experiments in this paper draw on multiple databases, summarized in Tab.~\ref{tab:datasets}. 
	BOSSBase (grayscale) and ALASKA2 (color) are popular in the steganography and steganalysis literature.
	From ImageNet, we use training set.
	The Dresden, VISION, and Forchheim databases are popular in the forensics community for benchmarking camera device or model identification. 
	Camera annotations make them ideal for camera-specific statistics and the study of prevalent compression patterns. 
	From the Dresden database, we use the natural scene subset. 
	From the Forchheim database, we use the camera-original subset. 
	Additionally, we acquired 201 flat-field frames with a Sony Alpha 6600 and a Nikon D850.

		\begin{table}[h]
		\caption{Databases used for illustrating the causes and effects of directionality}
		\label{tab:datasets}
		\begin{tabular}{lllr}
			\toprule
			Name & Purpose & Format & \# images \\
			\midrule
			BOSSBase~\cite{Bas2011Boss} & Steganalysis & NC & \numprint{10000} \\
			ALASKA2~\cite{Cogranne2020Alaska2} & Steganalysis & NC & \numprint{80005} \\
			ImageNet~\cite{Deng2009ImageNet} & Image recog. & JPEG & \numprint{1281141} \\
			Dresden~\cite{Gloe2010Dresden} & Forensics & JPEG & \numprint{16961} \\
			VISION (flat)~\cite{Shullani2017Vision} & Forensics & JPEG & \numprint{4167} \\
			VISION (nat)~\cite{Shullani2017Vision} & Forensics & JPEG & \numprint{7565} \\
			Forchheim~\cite{Hadwiger2020Forchheim} & Forensics & JPEG & \numprint{3851} \\
			Own (flat) & - & NC & \numprint{402} \\
			\bottomrule
			\multicolumn{4}{l}{NC = never compressed} 
		\end{tabular}
	\end{table}

\section{Comparison of directionality scores}
	\label{app:measuring_directionality}

	This section evaluates two methods to quantify the directionality of a given image. 
	The methods are based on the Sobel derivative filter and on steerable pyramids. 
	These two and related frequency-domain techniques have been proposed to characterize the distribution of orientation contours in natural images and textures~\cite{Coppola1998OrientationDistribution, Gorkani1994DominantOrientation, Hansen2004HorizontalBias, Duggan2023DigitalOrientationBias}. 
	We evaluate the ability of the methods to measure directionality from image transformations.

	\paragraph{Sobel} An intuitive approach to measuring directionality is to compute directional edges using a first-order derivative filter, such as the Sobel operator~\cite{Sobel2014GradientOperator}.  %
	The Sobel detector computes the derivative in horizontal and vertical directions. 
	A directionality score can be defined as the difference between the horizontal and vertical absolute-valued derivative image, normalized by their sum. 
	A positive score means that the image contains more vertical than horizontal edges.
	
	\paragraph{Steerable pyramids} 
	
	Steerable pyramids decompose an image's spectrum into multiple orientations and different frequency bands~\cite{Simoncelli1995SteerablePyramids}.
	After splitting the spectrum into higher and lower frequencies, the low-pass component is recursively subsampled and split into low--high and low--low frequencies. The recursive decomposition produces a pyramid, where each scale is decomposed into multiple orientation bands.
	
	For our directionality detector, we convert color images to grayscale and center-crop non-square images to a square resolution. The images are then decomposed into 16 orientation bands and $S = 2$ scales. 
	The choice of 16 bands ensures that the angular bands also include diagonal directional frequencies but contain little overlap between the horizontal and vertical regions. 
	We use $2$ scales as a compromise between including lower frequencies while discarding the lowest frequency components, as in a high-pass filter.
	The directionality score is the relative amount of energy in horizontal and vertical orientations bands, averaged over the two scales,
	
	\begin{equation}
		d = \frac{1}{S} \sum_{s=1}^{S} \frac{E_{s, h} - E_{s, v}}{E_{s, h} + E_{s, v}} \enspace.
	\end{equation}
	
	A positive score $d$ means that more energy is located in the horizontal frequencies; a negative score means that more energy is located in the vertical frequencies. 
	Note that horizontal frequencies correspond to vertical (top-to-bottom) edges.

	\begin{figure}
		\includegraphics{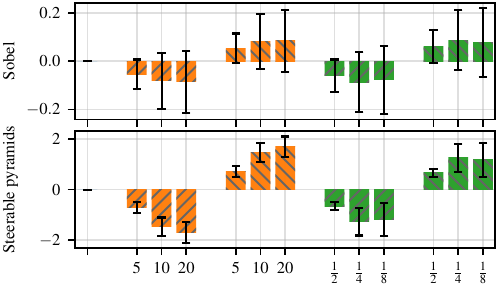}
		\caption{Comparison of directionality detectors: Sobel filters~(top) and steerable pyramids (bottom). Five types of test data:  symmetric synthetic images (leftmost), directionality added by horizontal/vertical smoothing (orange), and by horizontal/vertical downscaling (green). 
			Both detectors indicate the dominant image direction, but the steerable pyramids are more precise, as indicated by the tighter error bars.
		}
		\label{fig:directionality_detector_comparison_synthetic}
	\end{figure}
	
	\medskip %

	To compare the two methods, we start with \numprint{1000} \emph{symmetric} synthetic images of size $512^2$, consisting of 
	of cosine patterns with randomized amplitude %
	and additive Gaussian noise. %
	To add directionality in a controlled manner, we process these images with one out of four anisotropic transformations. 
	The first two are horizontal and vertical smoothing with a Hanning window of size $k \in [5, 10, 20]$~\cite[Ch.~7.2.1]{Oppenheim2009SignalProcessing}.
	The other two are horizontal or vertical downscaling by a scale factor $s \in \left[\frac{1}{2}, \frac{1}{4}, \frac{1}{8}\right]$, with Lanczos interpolation~\cite{Duchon1979Lanczos},
	and subsequent upscaling to the original resolution.

	Figure~\ref{fig:directionality_detector_comparison_synthetic} shows the average directionality scores of the Sobel~(top) and the steerable pyramids directionality detectors~(bottom). The error bars denote the standard deviation.
	As expected, both detectors assign a directionality score of $0$ to all unprocessed symmetric images. 
	Horizontal smoothing leads to a negative score because it attenuates horizontal frequencies. 
	So does horizontal downscaling.
	Both types of vertical transformations lead to positive scores.
	Overall, the error bars show that the steerable pyramids detector is more precise than the Sobel-based detector.
	
	We also evaluated the two directionality detectors with \numprint{1000} images from the ALASKA2 dataset~\cite{Cogranne2020Alaska2}. 
	All images were rotated by a random multiple of 90 degrees to remove directionality bias in the dataset.
	The results are similar to Fig.~\ref{fig:directionality_detector_comparison_synthetic}. %
	Since the individual original images already vary in their directionality (no real image is perfectly symmetric), the standard deviations are significantly larger, but the steerable pyramids remain more precise than the Sobel-based detector.
\section{CNN-based camera model identification}
	\label{app:cnn_based_camera_model_identification}
	
	This part has been moved to the appendix to make space for the SPAM feature descriptor, which allows us to directly measure the ``stress'' caused by explicit symmetrization.
	The headline result in Fig.~\ref{fig:introduction} use a CNN-based method for better comparability across applications.
	Here we describe the experiment and report on the effect of rotation augmentation.
	
	We replicate the setup of~\cite{Hadwiger2020Forchheim}, who train an EfficentNet-B5 as 27-class classifier on the Forchheim dataset, and pay particular attention to factors that affect directionality.
	We verified that 22 cameras stored their images in canonical orientation and set the EXIF orientation flag (cf.~Sec.~\ref{par:losslessrot}). The other cameras stored rotated images without orientation metadata. We use all images as stored to preserve the compression direction.
	
	Each image was subdivided into non-overlapping patches of size $256^2$. 
	From each image, we select the $100$ patches with the highest quality according to a heuristic quality metric~\cite{Bondi2017CamIdCnn}.
	This approach discards non-informative regions, e.g., consisting of homogeneous pixels.
	We trained on randomly cropped sub-patches of size $64^2$, using an initial learning rate of $10^{-5}$, and a dropout rate of $0.25$.
	The learning rate halves whenever the validation loss stagnates for 500 epochs.
	Training ends when the validation loss does not improve for 1000 epochs.
	
	We evaluate two setups. 
	In the \emph{no-rot} setup, the training patches are randomly flipped along the vertical and horizontal axis. 
	In the \emph{aug-rot} setup, the training patches are additionally rotated by a random multiple of 90 degrees, as in~\cite{Hadwiger2020Forchheim}.
	
	\begin{table}
		\caption{A CNN multi-class classifier for forensic camera model identification cannot maintain its accuracy on rotated test images. Rotation augmentation overcomes this limitation.}
		\centering
		\begin{tabular}{lcc}
			\toprule
			Training setup & Accuracy original & Accuracy rotated \\
			\midrule
			\emph{no-rot} & 0.941 & 0.845 \\
			\emph{aug-rot} & 0.932 & 0.929 \\
			\bottomrule
		\end{tabular}
		\label{tab:cnn_based_camera_model_identification}
	\end{table}

	Table~\ref{tab:cnn_based_camera_model_identification} reports the results. 
	In the \emph{no-rot} setup, the CNN achieves an accuracy of $0.941$.
	The rightmost column reports the accuracy after rotating the test images by 90 degrees in counter-clockwise direction.
	The drop in accuracy by $9.6$\,\%-pts.\ shows that the images contain directional information, which the CNN has learned to exploit.
	In the \emph{aug-rot} setup, the CNN achieves a slightly lower accuracy of $0.932$ but generalizes to rotated test images.

\section{SDXL autoencoder vs.\ text-to-image}
\label{app:autoencoder_vs_prompt}

	\begin{figure}
		\includegraphics{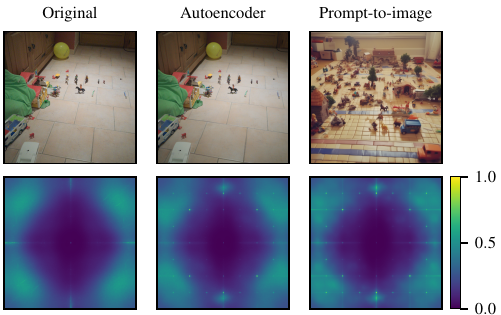}
		\caption{Passing a real image through SDXL's autoencoder inserts the same frequency-domain artifacts as in images generated from text prompts while preserving the scene content.}
		\label{fig:synthesis_artifacts}
	\end{figure}

	Today's synthetic image generators split the image generation into a semantic and a perceptual part~\cite{Rombach2022LatentDiffusion}.
	The perceptual part consists of an autoencoder that provides a compact image representation and can reconstruct images from this representation.
	The semantic part consists of a generative model that synthesizes the semantic content of the image in the autoencoder's compact representation.
	The autoencoder's decoder fills in the perceptual details while decompressing the generated representation to an image.

	We considered compressing the ALASKA2 dataset to text prompts with the CLIP interrogator\footnote{\url{https://github.com/pharmapsychotic/clip-interrogator/} with the CLIP model \textit{ViT-g-14/laion2B-s34B-b88K}.} and reconstructing a synthetic ALASKA2 dataset from the text prompts. 	
	Working with a small subset, we found that some of the generated images were surprisingly close to the original images.
	In many cases, however, the text prompts encouraged an artistic style or hallucinated animals. 
	We were concerned that this might introduce scene bias and discarded this approach.
		
	Our fake images in Sec.~\ref{sec:synthetic_image_detection} were created by passing real images through the generators' autoencoder.
	Figure~\ref{fig:synthesis_artifacts} demonstrates that this approach leaves the same artifacts as synthesizing images from text prompts.
	The top row shows a real image from ALASKA2 (left), the image after compression and decompression via the SDXL autoencoder (center), and an image synthesized from a text prompt (right).
	The bottom row shows the power spectrum after averaging \numprint{1000} noise residuals.
	The autoencoder image has the same scene content as the original image, but shows the same artifacts as images synthesized from text prompts.
	The artifacts in the right column appear slightly sharper because SDXL produces $1024^2$ images while the autoencoder maintains the $512^2$ image resolution.
	This shows that synthesis artifacts are introduced by the autoencoder, as also observed in related work~\cite{Cozzolino2023Clip}.
	
\section{Measurement of human contrast sensitivity}
	\label{app:contrast_sensitivity}
	
	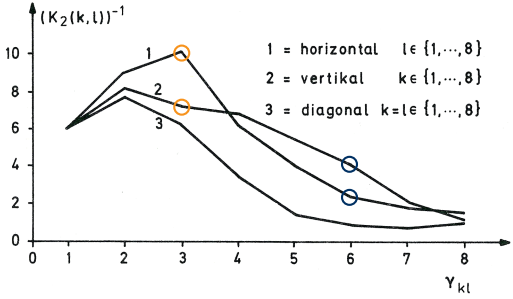
\begin{figure}
		\centering
		\begin{tikzpicture}
			\node[anchor=south west, inner sep=0] (image) at (0, 0) {\includegraphics[width=.8\columnwidth]{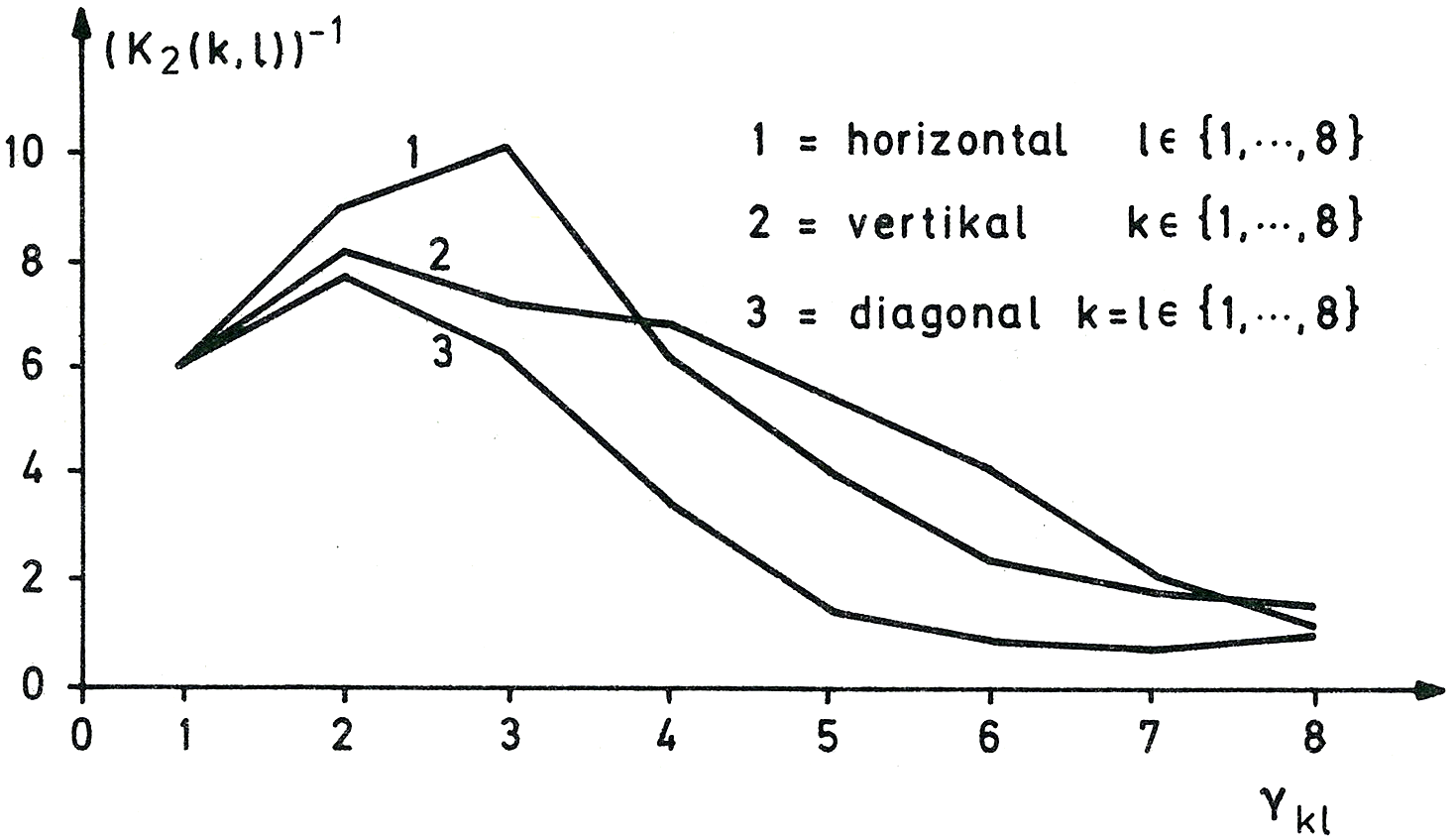}};
			\draw (2.37, 3.22) node [draw, circle, thick, maincolor, inner sep=2pt] {~};
			\draw (2.37, 2.5) node [draw, circle, thick, maincolor, inner sep=2pt] {~};

			\draw (4.6, 1.74) node [draw, circle, thick, altcolor, inner sep=2pt] {~};
			\draw (4.6, 1.3) node [draw, circle, thick, altcolor, inner sep=2pt] {~};
		\end{tikzpicture}
		\caption{Psychovisual thresholding experiments revealed different contrast sensitivity profiles for horizontal, vertical, and diagonal DCT basis functions. 
		Figure reproduced from~\cite[p.~138]{Lohscheller1982Diss}. 
		The circles are our own annotations, relating to the quantization factors highlighted in Fig.~\ref{fig:standard_qt}.}
		\label{fig:lohscheller_dct_contrast_sensitivity}
	\end{figure}
	
	Human contrast sensitivity to the 64 DCT basis functions was measured by Lohscheller in the early 1980s~\cite[pp.~122--142]{Lohscheller1982Diss}. 
	Figure~\ref{fig:lohscheller_dct_contrast_sensitivity} shows the measured contrast sensitivity for the horizontal, vertical, and diagonal DCT basis functions.
	The values are averages of 18 human subjects.
	Note that horizontal and vertical sensitivity are different and the order changes between low and high frequencies.
	These measurements are reflected in the standard JPEG quantization table (Fig.~\ref{fig:standard_qt}), which is still widely used today.

\section{Horizontal edges in human face perception}
	\label{app:stimuli}

\begin{figure}
		\includegraphics[width=\columnwidth]{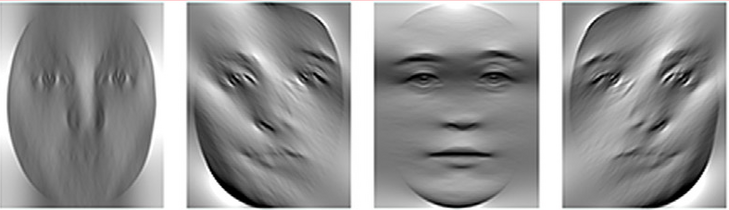}
		\caption{Orientation-filtered stimuli used in neuroscience. Images reproduced with kind permission of the authors~\cite{Goffaux2016Neuropsychologia}.}
		\label{fig:goffaux_stimuli} %
	\end{figure}

	Goffaux et al.\ confirmed that human brains prefer horizontal contours when recognizing human faces~\cite{Goffaux2016Neuropsychologia}.
	They presented orientation-filtered stimuli images to study participants in fMRI scanners. 
	These images are reproduced in Fig.~\ref{fig:goffaux_stimuli}.
	The strongest activity in high-level face-specialized ventral regions of the human brain was observed in the condition where horizontal edges were preserved, as in the third image.
	
\end{document}

%% file: figures/acquisition_pipeline.tex
	
	\begin{tikzpicture}[
		stage/.style={draw, rectangle, minimum width=2.5cm, minimum height=0.7cm, align=center},
		arrow/.style={-stealth},
		x=1cm,
		y=1cm,
		]
		
		\draw (1,0) node [stage] (scene) {Scene content\\{\small (Section~3.1)}};
		\draw (scene.east) node [right=1cm,stage] (sensor) {Sensor\\{\small (Section~3.2)}};
		\draw (sensor.east) node [right=1cm,stage] (raw-conversion) {Raw-to-image conversion\\{\small (Section~3.3)}};
		\draw (raw-conversion.east) node [right=1cm,stage] (compression) {Compression\\{\small (Section~3.4)}};
		
				
		\draw[arrow] (scene) -- (sensor);
		\draw[arrow] (sensor) -- (raw-conversion);
		\draw[arrow] (raw-conversion) -- (compression);
		
		
		\draw (scene) node [above=5mm] 	
		{\tikz[x=.75mm,y=.75mm]{
			\begin{scope}
				\clip [draw] (-2,-4.5) rectangle ++ (-16,9);
				\draw [fill,bottom color=white,top color=black!20] (-20,-1) rectangle (20,8);
				\draw [fill,top color=black!50,bottom color=black] (-20,-8) rectangle (20,-1);
			\end{scope}
			\begin{scope}
				\clip [draw] (2,-8) rectangle ++ (9,16);
				\draw [fill,bottom color=white,top color=black!20] (-20,-1) rectangle (20,8);
				\draw [fill,top color=black!50,bottom color=black] (-20,-8) rectangle (20,-1);
			\end{scope}
		}};

		\draw (sensor) node [above=5mm] 	
		{\tikz[x=.75mm,y=.75mm]{
			\begin{scope}
				\clip [draw] (-2,-4.5) rectangle ++ (-16,9);
				\foreach \x/\g in 
		{0.52/15,0.59/2,2.67/11,3.48/12,4.32/5,5.03/5,6.19/18,7.52/15,7.92/7,9.42/7,10.39/15,11.37/16,12.47/2,13.17/16,14.68/7,15.12/12,16.88/19,17.15/3,18.79/13,18.46/15,20.02/9,20.25/18,21.99/9,23.48/15,23.48/19,24.43/7,25.62/16,26.93/3,28.56/15,28.42/13,29.43/6,30.72/2,31.99/3,33.17/15,34.56/18,34.92/6,36.31/1,36.51/2,37.82/9,38.16/19,40.39/13}
					\draw [black!\g]  (-21,-4.5)++(\x,0)--++(0,9);
				\draw (-2,-4.5) rectangle ++ (-16,9);
			\end{scope}
			\begin{scope}
				\clip [draw] (2,-8) rectangle ++ (9,16);
				\foreach \y/\g in {0.52/15,0.59/2,2.67/11,3.48/12,4.32/5,5.03/5,6.19/18,7.52/15,7.92/7,9.42/7,10.39/15,11.37/16,12.47/2,13.17/16,14.68/7,15.12/12,16.88/19,17.15/3,18.79/13,18.46/15,20.02/9,20.25/18,21.99/9,23.48/15,23.48/19,24.43/7,25.62/16,26.93/3,28.56/15,28.42/13,29.43/6,30.72/2,31.99/3,33.17/15,34.56/18,34.92/6,36.31/1,36.51/2,37.82/9,38.16/19,40.39/13}
					\draw [black!\g]  (2,-8)++(0,\y)--++(9,0);
				\draw (2,-8) rectangle ++ (9,16);
			\end{scope}
		}};
		
		\draw (raw-conversion) node [above=5mm] 	
		{\tikz[x=.75mm,y=.75mm]{
			\begin{scope}
				\clip [draw] (-2,-4.5) rectangle ++ (-16,9);
				\draw [fill,black!20] (-20,-8) rectangle (20,8);
				\draw [fill,white] (-5.5,-4.5) rectangle ++(-9,9);
			\end{scope}
			\begin{scope}
				\clip [draw] (2,-8) rectangle ++ (9,16);
				\draw [fill,black!20] (-20,-8) rectangle (20,8);
				\draw [fill,white] (2,-4.5) rectangle ++(9,9);
			\end{scope}
		}};

		\draw (compression) node [above=5mm] 	
		{\tikz[x=.75mm,y=.75mm]{
			\begin{scope}
				\clip [draw] (-2,-4.5) rectangle ++ (-16,9);
				\draw [fill,black!20,xstep=1,ystep=.5,very thin] (-20,-8) grid (20,8);
				\draw (-2,-4.5) rectangle ++ (-16,9);
			\end{scope}
			\begin{scope}
				\clip [draw] (2,-8) rectangle ++(9,16);
				\draw [fill,black!20,xstep=1,ystep=.5,very thin] (-20,-8) grid (20,8);
				\draw (2,-8) rectangle ++(9,16);
			\end{scope}
			\begin{scope} 	
				\clip [draw] (13,-8) rectangle ++ (9,16);
				\draw [fill,black!20,xstep=.5,ystep=1,very thin] (-20,-8) grid (22,8);
				\draw (13,-8) rectangle ++ (9,16);
			\end{scope}
		}};

	\end{tikzpicture}
	